\documentclass{article}

\usepackage{graphicx}
\usepackage{dcolumn}
\usepackage{bm}
\usepackage[latin1]{inputenc}

\usepackage[english]{babel}
\usepackage{float}
\usepackage{makeidx}
\usepackage{amssymb}
\usepackage{amsfonts}
\usepackage{amsmath}
\usepackage{graphicx}

\begin{document}

\title{Quantum discrete breathers}
\author{Ricardo A. Pinto$^1$ and Sergej Flach$^2$
\\ \\
{\small 1. Department of Electrical Engineering, University of California Riverside,}\\
{\small 900 University Ave., Riverside, California 92521, USA}
\\
{\small 2. Max-Planck-Institut f\"ur Physik komplexer Systeme,}\\
{\small N\"othnitzer
Str. 38, 01187 Dresden, Germany
}}


\date{\today}




\maketitle


\section{Introduction}

In solid-state physics, the phenomenon of localization is usually perceived as arising from
extrinsic disorder that breaks the discrete translational invariance of the perfect
crystal lattice. Familiar examples include the localized vibrational modes around impurities
or defects in crystals and Anderson localization of waves in disordered media \cite{Anderson1958}.
The usual perception among solid-state researchers is that, in perfect lattices excitations
must be extended objects as well, essentially plane wave like. Such firmly entrenched
perceptions were severely jolted since the discovery of discrete breathers (DB), also 
known as intrinsic localized modes (ILM). These states are typical excitations in perfectly
periodic but strongly nonlinear systems, and are characterized by being spatially localized,
at variance to plane wave states \cite{FlachPhysRep295,CFK04,FlachPhysRep467}.

DB-like excitations, being generic objects, have been observed in a large
variety of lattice systems that include bond excitations in molecules, lattice
vibrations and spin excitations in solids, charge flow in coupled
Josephson junctions, light propagation in interacting optical waveguides,
cantilever vibrations in michromechanical arrays, cold atom dynamics in
Bose-Einstein condensates loaded on optical lattices, among others. They
have been extensively studied, and a high level of understanding about
their properties has been reached.

Two decades of intensive research have polished our theoretical understanding of DBs
in classical nonlinear lattices. Less is known about their quantum counterparts -
quantum breathers (QB). This chapter is devoted to a review of the more recent studies
in this field. The concept of QBs is closely related with the theme of
dynamical tunneling in phase space.

\subsection{A few facts about classical discrete breathers}

Let us study the combined effect of
nonlinearity and discreteness on the spatial localization of
a discrete breather on a basic level. 
For that we look into the dynamics
of a one-dimensional chain of interacting (scalar) oscillators 
with the Hamiltonian
\begin{equation}
H = \sum_n \left[ \frac{1}{2}p_n^2 + V(x_n) + W(x_n - x_{n-1}) \right]\;.
\label{1.2-1}
\end{equation}
The integer $n$ marks the lattice site number of a possibly
infinite chain, and $x_n$ and $p_n$ are the canonically conjugated
coordinate and momentum of a degree of freedom associated with site number $n$.
The on-site potential $V$ and the interaction potential $W$
satisfy $V'(0)=W'(0)=0$,
$V''(0),W''(0) \geq 0$.
This choice ensures that the classical ground state $x_n=p_n=0$
is a minimum of the energy $H$.
The equations of motion read
\begin{equation}
\dot{x}_n=p_n\;,\;\dot{p}_n=-V'(x_n) -W'(x_n - x_{n-1})
+ W'(x_{n+1}-x_n)\;. 
\label{1.2-2}
\end{equation}

Let us linearize the equations of motion around the classical ground state.
We obtain a set of linear coupled differential equations
with solutions being small amplitude plane waves:
\begin{equation}
x_n(t) \sim {\rm e}^{{\rm i}(\omega_q t - qn)} \;,\;
\omega_q^2 = V''(0) + 4W''(0)\sin^2 \left(\frac{q}{2}\right)\;.
\label{1.2-3}
\end{equation}
These waves are characterized by a wave number $q$ and
a corresponding frequency $\omega_q$. All allowed plane
wave frequencies fill a part of the real axis which
is coined linear spectrum. Due to the underlying lattice
the frequency $\omega_q$ depends periodically on $q$ and
its absolute value has always a {\it finite upper bound}.
The maximum (Debye) frequency of small amplitude waves  
$\omega_{\pi}=\sqrt{V''(0) + 4W''(0)}$.
Depending on the choice of the potential $V(x)$, $\omega_q$ can be either acoustic- or optic-like,
$V(0)=0$ and $V(0)\ne 0$, respectively. 
In the first case the linear spectrum covers the interval
$ -\omega_{\pi} \leq \omega_q \leq \omega_{\pi}$ which includes
$\omega_{q=0}=0$. 
In the latter case an additional
(finite) gap opens for $|\omega_q|$
below the value $\omega_{0}=\sqrt{V''(0)}$.

For large amplitude excitations the linearization
of the equations of motion is not correct anymore.
Similar to the case of a single anharmonic oscillator,
the frequency of possible time-periodic excitations
will depend on the amplitude of the excitation,
and thus may be located outside the linear spectrum.
Let us assume that a time-periodic
and spatially localized state, 
i.e. a \emph{discrete breather}, $\hat{x}_n(t+T_b)=\hat{x}_n(t)$ 
exists as an exact solution of Eqs.(\ref{1.2-2}) with the period
$T_b=2\pi/\Omega_b$. 
Due to its time periodicity, we can
expand $\hat{x}_n(t)$ into a Fourier series
\begin{equation}
\hat{x}_n(t) = \sum_k A_{kn}{\rm e}^{ik\Omega_b t}\;.
\label{1.2-4}
\end{equation}
The Fourier coefficients are by assumption also localized
in space
\begin{equation}
A_{k,|n| \to \infty} \to 0 \;.
\label{1.2-5}
\end{equation}
Inserting this ansatz into the equations of motion (\ref{1.2-2}) and
linearizing the resulting algebraic equations for Fourier coefficients in
the spatial 
breather tails (where the amplitudes are by assumption small)
we arrive at the
following linear algebraic equations:
\begin{equation}
k^2 \Omega_b^2 A_{kn} = V''(0) A_{kn} + W''(0) (2A_{kn} - A_{k,n-1}
- A_{k,n+1})
\;.
\label{1.2-6}
\end{equation}
If $k\Omega_b = \omega_q$,
the solution to (\ref{1.2-6}) is 
$A_{k,n} = c_1 {\rm e}^{iqn} + c_2 {\rm e}^{-iqn}$.
Any nonzero (whatever small) amplitude $A_{k,n}$ wil thus
oscillate without further spatial decay, contradicting
the initial assumption. 
If however
\begin{equation}
k \Omega_b \neq \omega_q
\label{1.2-7}
\end{equation}
for any integer $k$ and any $q$, then the general solution to (\ref{1.2-6})
is given by $A_{k,n} = c_1 \kappa^n + c_2 \kappa^{-n}$ where
$\kappa$ is a real number depending on $\omega_q$, $\Omega_b$ and $k$.
It always admits an (actually exponential) spatial
decay by choosing either $c_1$ or $c_2$ to be nonzero.
In order to fulfill (\ref{1.2-7}) for at least one real value
of $\Omega_b$ and any integer $k$, we have to request $|\omega_q|$
to be bounded from above. That is precisely the reason why 
the spatial lattice is needed. In contrast most spatially
continuous field equations will have linear spectra which are
unbounded. That makes resonances
of higher order harmonics of a localized excitation with the linear
spectrum unavoidable.
The nonresonance condition (\ref{1.2-7}) is thus an (almost) necessary
condition for obtaining a time-periodic localized state
on a Hamiltonian lattice 
\cite{FlachPhysRep295}. 

The performed analysis can be extended to more general classes of discrete
lattices, including e.g. long-range interactions between sites, more
degrees of freedom per each site, higher-dimensional lattices etc.
But the resulting non-resonance condition (\ref{1.2-7}) keeps its
generality, illustrating the key role of discreteness and nonlinearity for the
existence of discrete breathers.

\begin{figure}[!t]
\centering
\includegraphics[width=0.5\textwidth]{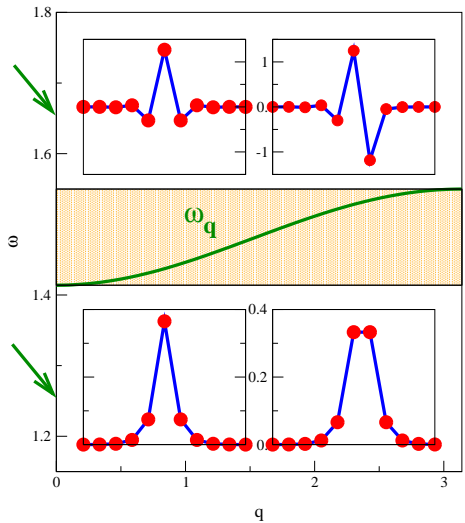}
\caption{\label{fig1.4-1}The frequency versus wavenumber dependence
of the linear spectrum for a one-dimensional
chain of anharmonic oscillators with potentials (\ref{1.4-1}). 
Chosen DB frequencies are marked
with green arrows and lie outside the linear spectrum $\omega_q$.
Red circles indicate the oscillator displacements for a given
DB solution, with all velocities equal to zero. Lines connecting
circles are guides for the eye. From \cite{CFK04}.}
\end{figure}

Let us show discrete breather solutions for various lattices.
We start with a chain (\ref{1.2-1}) with the
functions
\begin{equation}
V(x)=x^2+x^3+\frac{1}{4}x^4\;,\;W(x)=0.1 x^2\;.
\label{1.4-1}
\end{equation}
The spectrum $\omega_q$ is optic-like and shown in Fig.\ref{fig1.4-1}.
Discrete breather solutions can have frequencies $\Omega_b$ which
are located both below and above the linear spectrum.
The time-reversal symmetry of (\ref{1.2-2}) allows to
search for DB displacements $x_n(t=0)$ when all velocities
$\dot{x}_n(t=0)=0$. These initial displacements are computed
with high accuracy (see following sections) and plotted
in the insets in Fig.\ref{fig1.4-1} \cite{CFK04}. 
We show solutions
to two DB frequencies located above and below $\omega_q$ -
their actual values are marked with the green arrows.
To each DB frequency we show two different spatial DB patterns -
among an infinite number of other possibilities, as we will see below.
The high-frequency DBs ($\Omega_b \approx 1.66$) occur for large-amplitude,
high-energy motion with adjacent particles moving out of phase.
Low-frequency DBs ($\Omega_b \approx 1.26$) occur
for small-amplitude motion with adjacent particles
moving in phase.

\begin{figure}[!t]
\centering
\includegraphics[angle=-90,width=0.5\textwidth]{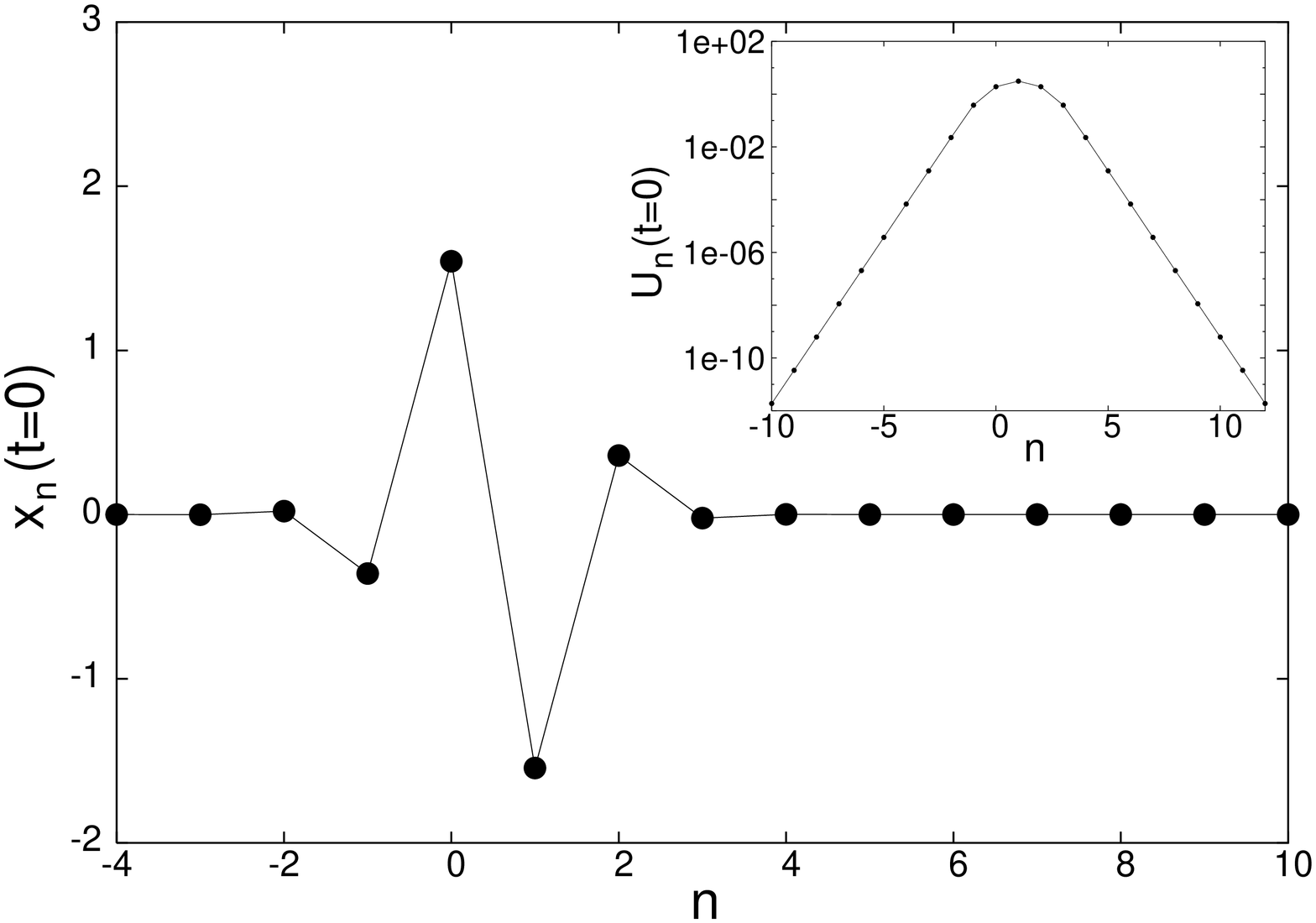}
\includegraphics[angle=-90,width=0.5\textwidth]{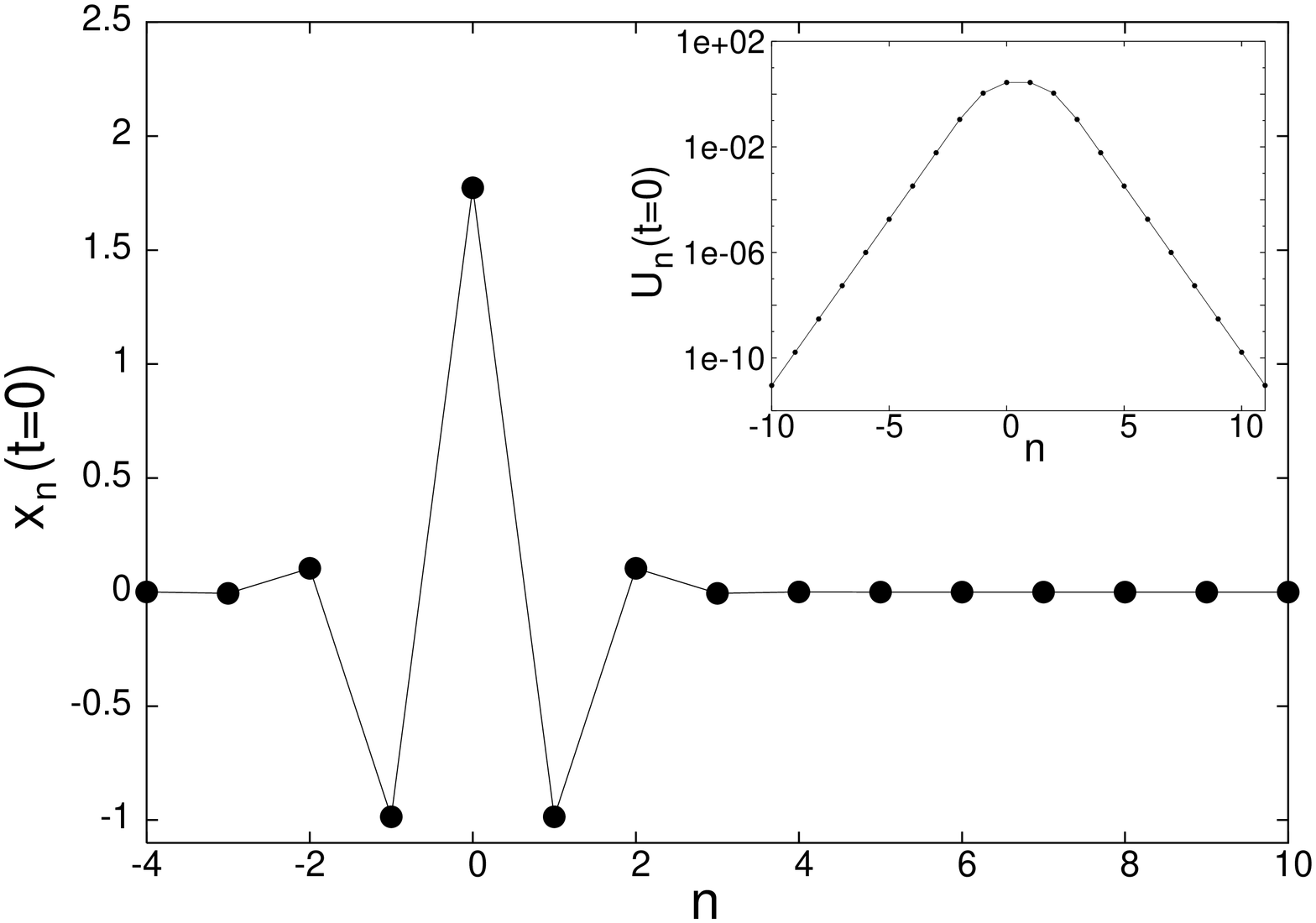}
\caption{\label{fig1.4-3}Discrete breather solutions for a Fermi-Pasta-Ulam chain (see text). 
These states are frequently referred to as the Page mode (left) and
the Sievers-Takeno mode (right).
Adapted from \cite{fg05chaos}.}
\end{figure}

In Fig.\ref{fig1.4-3} we show two DB solutions for a Fermi-Pasta-Ulam
chain of particles coupled via anharmonic springs $V(x)=0,
W(x)=\frac{1}{2}x^2+\frac{1}{4}x^4$ (c.f. (\ref{1.2-1}))
which has an acoustic-type spectrum \cite{fg05chaos}.
The DB frequency is in both cases $\Omega_b=4.5$.
Again the displacements $x_n$ are shown for an initial time when
all velocities vanish. In the inset we plot the strain
$u_n=x_n-x_{n-1}$ on a log-normal scale. The DB solutions
are exponentially localized in space.

\begin{figure}[!t]
\centering
\includegraphics[angle=0,width=0.9\textwidth]{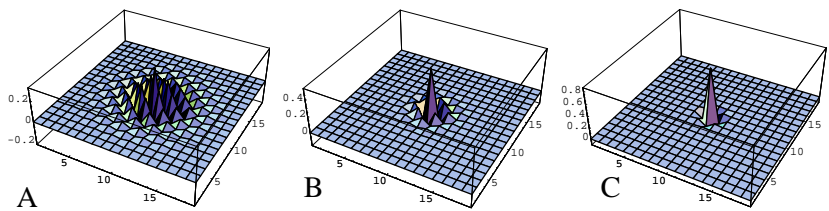}
\caption{\label{fig1.4-2}Displacements of DBs on a two-dimensional lattice 
(\ref{1.4-2}) with $k=0.05$, all velocities equal to zero. (A) $\Omega_b=1.188$;
(B) $\Omega_b=1.207$; (C) $\Omega_b=1.319$. From \cite{ef05pd}.}
\end{figure}

Finally we show DB solutions for a {\it two-dimensional}
square lattice of anharmonic oscillators with nearest
neighbour coupling. The equations of motion read
\begin{equation}
\ddot x_{i,j}=k(x_{i+1,j}+x_{i-1,j}-2x_{i,j}) +k(x_{i,j+1}+x_{i,j-1}-2x_{i,j})
-x_{i,j}-x_{i,j}^3
\label{1.4-2}
\end{equation}
with oscillator potentials $V(x)=\frac{1}{2}x^2+
\frac{1}{4}x^4$. In Fig.\ref{fig1.4-2} we plot the 
oscillator displacements with all velocities equal to zero for
three different DB frequencies and $k=0.05$ 
\cite{ef05pd}.
For all cases adjacent oscillators move out of phase.

We conclude this section with emphasizing that DB solutions can be
typically localized on a few lattice sites, regardless of the 
lattice dimension. Thus little overall coherence is needed to excite
a state nearby - just a few sites have to oscillate coherently,
the rest of the lattice does not participate strongly in the excitation.

\subsection{From classical to quantum}

A natural question is what remains of discrete breathers
if the corresponding quantum problem is considered \cite{Aubry1997PhysicaD103,rsm00pa,Fleurov2003}. 
The many-body Schr{\"o}dinger equation is linear and translationally invariant, therefore 
all eigenstates must obey the Bloch theorem. Thus
we cannot expect eigenstates of the Hamiltonian to be 
spatially localized (on the lattice). What is the
correspondence between the quantum eigenvalue problem and
the classical dynamical evolution?

The concept of tunneling is a possible 
answer to this puzzle. Naively 
speaking we quantize the family of periodic orbits associated with a
discrete breather located somewhere on the lattice. Notice that
there are as many such families as there are lattice sites.
The quantization (e.g., Bohr-Sommerfeld) yields some
eigenvalues. Since we can perform the same procedure with
any family of discrete breather periodic orbits which differ
only in their location on the lattice, we obtain $N$-fold
degeneracy for every thus obtained eigenvalue, where $N$ stands
for the number of lattice sites. Unless we consider the trivial
case of uncoupled lattice sites, these degeneracies will
be lifted. Consequently, we will instead obtain bands of states
with finite band width. 
These bands will be called quantum breather
bands. The inverse tunneling 
time of a semiclassical breather from one site to a neighboring one is
a measure of the bandwidth. 

We can then formulate the following expectation: if a classical nonlinear 
Hamiltonian lattice possesses discrete breathers, its quantum counterpart 
should show up with nearly degenerate bands of eigenstates, if the classical 
limit is considered. The number of states in such a band is $N$, and the 
eigenfunctions are given by Bloch-like superpositions of the semiclassical 
eigenfunctions obtained using the mentioned Bohr-Sommerfeld quantization of 
the classical periodic orbits. By nearly degenerate we mean that 
the bandwidth of a quantum breather band is much smaller than 
the spacing 
between different breather bands and the average level spacing in
the given energy domain, and the 
classical limit implies large eigenvalues. 

Another property of a quantum breather state is that such
a state shows up with exponential localization in appropriate 
correlation functions \cite{wgbs96}.
This approach selects all many-particle bound
states, no matter how deep one is in the quantum regime.
In this sense quantum breather states belong to the class of
many-particle bound states.

For large energies and $N$ the
density of states becomes large too. What will happen to the
expected quantum breather bands then? Will the hybridization
with other non-breather states destroy the particle-like nature
of the quantum breather, or not? What is the impact of the
nonintegrability of most systems allowing for classical 
breather solutions? Since the quantum case corresponds to
a quantization of the classical phase space, we could expect
that chaotic trajectories lying nearby classical breather solutions
might affect the corresponding quantum eigenstates. 

From a computational point of view we are very much restricted
in our abilities to study quantum breathers. Ideally we would
like to study quantum properties of a lattice problem in
the high energy domain (to make contact with classical states)
and for large lattices. This is typically impossible, since
solving the quantum problem amounts to diagonalizing the Hamiltonian
matrix with rank $b^N$ where $b$ is the number of states per site,
which should be large to make contact with classical dynamics.
Thus typically quantum breather states 
have been so far obtained numerically for small 
one-dimensional systems \cite{wgbs96,sasrewpgw96,pdmacsjcjce91}.

\section{Quantum breather models}

\subsection{The Bose-Hubbard chain}
\label{sec7.1}

Let us discuss quantum breathers within the widely used 
quantum discrete 
nonlinear Schr{\"o}dinger model (also called
Bose-Hubbard model) with the Hamiltonian \cite{acsjcehg94}
\begin{equation}
H = -\sum_{l=1}^N\left[ \frac{1}{2} a_l^{\dagger}a_l^{\dagger}a_la_l
+ C(a_l^{\dagger}a_{l+1} + h.c.) \right]
\label{7-1}
\end{equation}
and the commutation relations
\begin{equation}
a_la^{\dagger}_m - a^{\dagger}_m a_l = \delta_{lm}
\label{7-2}
\end{equation}
with $\delta_{lm}$ being the standard Kronecker symbol.
This Hamiltonian conserves the total number of particles 
\begin{equation}
B = \sum_l n_l\;,\; n_l = a_l^{\dagger}a_l \;.
\label{7-3}
\end{equation}
For $b$ particles and $N$ sites the number of basis states
is 
\begin{equation}
\frac{(b+N-1)!}{b! (N-1)!}
\;.
\label{7-4}
\end{equation}
For $b=0$ there is just one trivial state of an empty lattice.
For $b=1$ there are $N$ states which correspond to one-boson
excitations. These states are similar to classical
extended wave states. For $b=2$ the problem is still exactly
solvable, because it corresponds to a two-body problem on a lattice.
\begin{figure}[!t]
\centering
\includegraphics[angle=-90,width=0.5\textwidth]{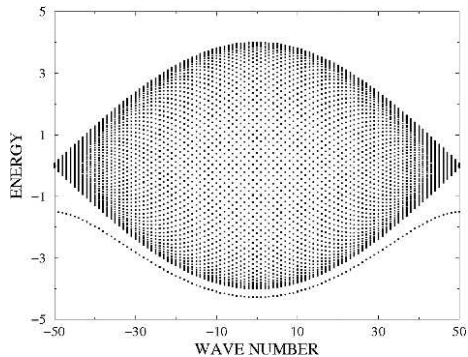}
\caption{\label{fig7-1}Spectrum of the quantum DNLS with $b=2$ and $N=101$.
The energy eigenvalues are plotted versus the wavenumber of
the eigenstate. Adapted from \cite{acsjcehg94}.}
\end{figure}
A corresponding numerical solution is sketched in Fig.~\ref{fig7-1}.
Note the wide two-particle continuum, and a single band located
below. This single band corresponds to quasiparticle states
characterized by one single quantum number (related to the wavenumber $q$).
These states are two-particle bound states.
The dispersion of this band is given \cite{acsjcehg94} by 
\begin{equation}
E = -\sqrt{1 + 16C^2 \cos^2 \left(\frac{q}{2}\right) }
\;.
\label{7-5}
\end{equation}
Any eigenstate from this two-particle bound state band is
characterized by exponential localization of correlations,
i.e. when represented in some set of basis states, the
amplitude or overlap with a basis state where the two particles
are separated by some number of sites is exponentially decreasing
with increasing separation distance. Note that a compact
bound state is obtained for $q=\pm \pi$, i.e. for these
wave numbers basis states with nonzero separation distance do
not contribute to the eigenstate at all \cite{jce03}. 

\subsection{The dimer}
\label{sec7.2}

A series of papers was devoted to the properties of the quantum
dimer \cite{bes90,b93pd,afko96prl,gkarb02}.
This system  
describes the dynamics of bosons fluctuating between two
sites. The number of bosons is conserved, and together with
the conservation of energy the system appears to be integrable.
Of course, one cannot consider spatial localization in such
a model. However, a reduced form of the discrete translational
symmetry - namely the permutational symmetry of the two sites -
can be imposed. Together with the addition of nonlinear terms
in the classical equations of motion the dimer allows for
classical trajectories which are not invariant under permutation.
The phase space can be completely analyzed, all isolated
periodic orbits (IPO) can be found. There appears exactly one 
bifurcation 
on one family of isolated periodic orbits, which 
leads to the appearance of a separatrix in phase space. The separatrix
separates three regions - one invariant and two non-invariant
under permutations. The subsequent analysis of the quantum
dimer demonstrated the existence of pairs of eigenstates with
nearly equal eigenenergies \cite{bes90}.
The separatrix and the bifurcation
in the classical phase space can be traced in the spectrum
of the quantum dimer \cite{afko96prl}.

The classical Hamiltonian may be written as
\begin{equation}
H = \Psi^*_1\Psi_1 + \Psi^*_2\Psi_2 + \frac 12 \left( (\Psi^*_1\Psi_1)^2
+ (\Psi^*_2\Psi_2)^2 \right) + C \left( \Psi^*_1\Psi_2 + \Psi^*_2\Psi_1
\right) \;\;. \label{7-2-3}
\end{equation}
with the equations of motion $\dot{\Psi}_{1,2} = i\partial H / \partial
\Psi^*_{1,2}$.
The model conserves the norm (or number of particles) 
$B=|\Psi_1|^2 + |\Psi_2|^2$.

Let us parameterize the phase space of 
(\ref{7-2-3}) with $\Psi_{1,2}=A_{1,2}{\rm e}^{i\phi_{1,2}}$, 
$A_{1,2} \geq 0$. It follows that $A_{1,2}$ is time independent and 
$\phi_1=\phi_2 + \Delta$ with $\Delta=0,\pi$ and $\dot{\phi}_{1,2}=\omega$ 
being also time independent. Solving the algebraic equations
for the amplitudes of the IPOs we obtain
\begin{eqnarray}
{\rm I:}\; A_{1,2}^2 = \frac 12 B \;,\; \Delta=0\;,\;
\omega=1+C+\frac 12 B \;\;,\label{7-2-4a} \\
{\rm II:}\; A_{1,2}^2 = \frac 12 B \;,\; \Delta=\pi\;,\;
\omega=1-C+\frac 12 B \;\;,\label{7-2-4b} \\
{\rm III:}\; A_1^2=\frac 12 B\left(1\pm \sqrt{1-4C^2/B^2}\right)\;,\;
\Delta=0\;,\;\omega = 1+B\;\;.\label{7-2-4c}
\end{eqnarray}
IPO III corresponds to two elliptic solutions which break the permutational
symmetry. IPO III exists for $B \geq B_b$ with $B_b=2C$ and occurs through
a bifurcation from IPO I.
The corresponding separatrix manifold is uniquely
defined by the energy of IPO I at a given value of $B \geq B_b$. This manifold
separates three regions in phase space - two with symmetry broken solutions,
each one containing one of the IPOs III, and one with symmetry conserving
solutions containing the elliptic IPO II. The separatrix manifold
itself contains the
hyperbolic IPO I. For $B \leq B_b$ only two IPOs exist - IPO I and II, with
both of them being of elliptic character. Remarkably there exist no other
IPOs, and the mentioned bifurcation and separatrix manifolds are the only
ones present in the classical phase space of (\ref{7-2-3}).

To conclude the analysis of the classical part, we list the energy
properties of the different phase space parts separated by the separatrix
manifold. First it is straightforward to show that the IPOs
(\ref{7-2-4a})-(\ref{7-2-4c}) correspond to maxima, minima or saddle points
of the energy in the allowed energy interval for a given value of $B$, with
no other extrema or saddle points present. It follows
\begin{eqnarray}
E_1=H({\rm IPO\; I}) = B + \frac 14 B^2 + CB \;\;,\label{7-2-5a} \\
E_2=H({\rm IPO\; II}) = B + \frac 14 B^2 - CB \;\;,\label{7-2-5b} \\
E_3=H({\rm IPO\; III}) = B + \frac 12 B^2 + C^2 \;\;. \label{7-2-5c}
\end{eqnarray}
For $B < B_b$ we have $E_1 > E_2$ (IPO I - maximum, IPO II - minimum). For
$B \geq B_b$ it follows $E_3 > E_1 > E_2$ (IPO III - maxima, IPO I - saddle,
IPO II - minimum). If $B < B_b$, then all trajectories are symmetry conserving.
If $B \geq B_b$, then trajectories with energies $E_1 < E \leq E_3 $ are
symmetry breaking, and trajectories with $E_2 \leq E \leq E_1$ are symmetry
conserving.

The quantum eigenvalue problem 
amounts to replacing the complex functions $\Psi, \Psi^*$
in (\ref{7-2-3}) by the boson annihilation and creation operators
$a,a^{\dagger}$ with the standard commutation relations
(to enforce the invariance under the exchange 
$\Psi \Leftrightarrow \Psi^*$ the substitution has to be
done on rewriting
$\Psi\Psi^*=1/2(\Psi \Psi^* + \Psi^* \Psi)$):
\begin{equation}
H = \frac{5}{4} + \frac{3}{2}\left(a_1^{\dagger}a_1 + a_2^{\dagger}a_2 \right)
+ \frac 12 \left( (a_1^{\dagger}a_1)^2 + (a_2^{\dagger}a_2)^2 \right)
+ C \left( a_1^{\dagger}a_2 + a_2^{\dagger}a_1 \right) \;\;. \label{7-2-6}
\end{equation}
Note that $\hbar = 1$ here, and the eigenvalues $b$ of $B=a_1^{\dagger}a_1 + a_2^{\dagger}a_2$
are integers. Since $B$ commutes with $H$ we can diagonalize the
Hamiltonian in the basis of eigenfunctions of $B$. Each value of $b$ 
corresponds to a subspace of the 
dimension $(b+1)$ in the space of eigenfunctions 
of $B$. These eigenfunctions are products of the 
number states $|n\rangle$ of each 
degree of freedom and can be characterized by a 
symbol $|n,m\rangle$ with $n$ bosographicsns in
the site 1 and $m$ bosons in the site 2. 
For a given value of $b$ it follows $m=b-n$.
So we can actually label each state by just one 
number $n$: $|n,(b-n)\rangle \equiv |n)$.
Consequently the eigenvalue problem at fixed $b$ amounts to diagonalizing
the matrix
\begin{equation}
H_{nm}= \left\{
\begin{array}{lr}
\frac{5}{4} + \frac{3}{2}b + \frac{1}{2}\left( n^2 + (b-n)^2
\right) & n=m \\
C\sqrt{n(b+1-n)} & n=m+1 \\
C\sqrt{(n+1)(b-n)} & n=m-1 \\
0                  & {\rm else}
\end{array}
\right
. \label{7-2-8}
\end{equation}
where $n,m=0,1,2,...,b$. Notice that the matrix $H_{nm}$ is a symmetric
band matrix. The additional symmetry $H_{nm}=H_{(b-n),(b-m)}$ is a
consequence of the permutational symmetry of $H$.
\begin{figure}[!t]
\centering
\includegraphics[angle=-90,width=0.5\textwidth]{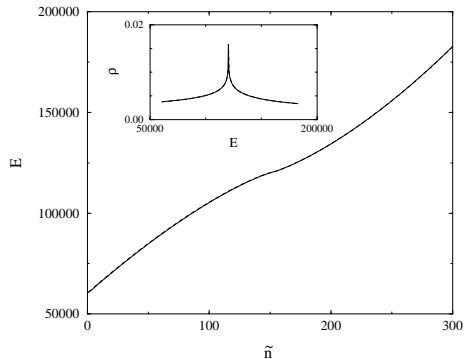}
\caption{\label{fig7-2}Eigenvalues versus ordered state number $\tilde{n}$
for symmetric and antisymmetric states  ($0 < \tilde{n} < b/2$
for both types of states). Parameters: $b=600$ and $C=50$.
Inset: Density of states versus energy.   
Adapted from \cite{afko96prl}.}
\end{figure}
For $C=0$ the matrix $H_{nm}$ is diagonal, with the property that
each eigenvalue is doubly degenerate (except for the state $|b/2)$ for
even values of $b$). The classical phase space contains only symmetry broken
trajectories, with the exception of IPO II and the separatrix with IPO I (in
fact in this limit the separatrix manifold is nothing but a resonant torus
containing both IPOs I and II). With the exception of the separatrix
manifold, all tori break permutational symmetry and come in two groups
separated by the separatrix. Then quantizing each group will lead to pairs
of degenerate eigenvalues - one from each group. There is a clear
correspondence to the spectrum of the diagonal ($C=0$) matrix 
$H_{nm}$. The eigenvalues $H_{00}=H_{bb}$ correspond to the quantized IPOs
III. With increasing $n$ the eigenvalues $H_{nn}=H_{(b-n),(b-n)}$ correspond
to quantized tori further away from the IPO III. Finally the states
with $n=b/2$ for even $b$ or $n=(b-1)/2$ for odd $b$ are tori most close
to the separatrix. Switching the side diagonals on by increasing $C$ will
lead to a splitting of all pairs of eigenvalues. In the case of small values
of $b$ these splittings have no correspondence to classical system properties.
However, in the limit of large $b$ we enter the semiclassical regime, and due
to the integrability of the system, eigenfunctions should correspond to tori
in the classical phase space which satisfy the Einstein-Brillouin-Keller
quantization rules.
\begin{figure}[!t]
\centering
\includegraphics[angle=-90,width=0.5\textwidth]{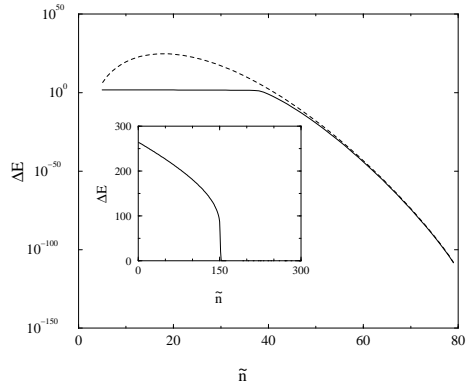}             
\caption{\label{fig7-3}Eigenvalue splittings versus $\tilde{n}$ 
for $b=150$ and $C=10$. Solid line - numerical result,
dashed line - perturbation theory. 
Inset: Same for $b=600$ and $C=50$. Only numerical results
are shown. Adapted from \cite{afko96prl}.}
\end{figure}
Increasing $C$ from zero will lead to a splitting $\Delta E_n$ of the 
eigenvalue doublets of $C=0$. 
In other words, we find pairs of eigenvalues, which
are related to each other through the symmetry of their eigenvectors and
(for small enough $C$) through the small value of the splitting.
These splittings have been calculated numerically and using perturbation
theory \cite{bes90,afko96prl}.
In the limit of large $b$ the splittings are exponentially
small for energies above the classical separatrix energy (i.e. for 
classical trajectories which are not invariant under permutation). 
If the eigenenergies are lowered below the classical separatrix energy,
the splittings grow rapidly up to the mean level spacing.

In Fig.~\ref{fig7-2} the results of a diagonalization of a system with
600 particles ($b=600$) is shown \cite{afko96prl}.
The inset shows the density of
states versus energy, which nicely confirms the predicted singularity
at the energy of the separatrix of the classical counterpart.
In order to compute the exponentially small splittings,
we may use e.g. a Mathematica routine which allows to choose
arbitrary values for the precision of computations. Here we chose
precision 512.
In Fig.~\ref{fig7-3} the numerically computed splittings are compared
to perturbation theory results. As expected, the splittings become
extremely small above the separatrix. Consequently these states
will follow for long times the dynamics of a classical broken symmetry
state.

\subsection{The trimer }
\label{sec7.3}

The integrability of the dimer does not allow a study
of the influence of chaos (i.e. nonintegrability) on the tunneling properties of
the mentioned pairs of eigenstates. A natural extension of
the dimer to a trimer adds a third degree of freedom without
adding a new integral of motion. Consequently the trimer
is nonintegrable. 
A still comparatively simple numerical
quantization of the trimer allows to study the behavior of 
many tunneling states
in the large-energy domain of the eigenvalue spectrum \cite{ff97jpc}.

Similarly to the dimer, the quantum trimer Hamiltonian is represented in
the form
\begin{eqnarray}
H=\frac{15}{8} + \frac{3}{2}(a_1^{\dagger}a_1 + a_2^{\dagger}a_2 + a_3^{\dagger}a_3) +
\frac{1}{2}\left[ (a_1^{\dagger}a_1)^2 + (a_2^{\dagger}a_2)^2 \right] 
\nonumber \\
+ C(a_1^{\dagger}a_2 + a_2^{\dagger}a_1) + \delta (a_1^{\dagger}a_3 + a_3^{\dagger}a_1 +
a_2^{\dagger}a_3 + a_3^{\dagger}a_2)\;\;.\label{7-3-7}   
\end{eqnarray}
Again $B=a_1^{\dagger}a_1 + a_2^{\dagger}a_2 + a_3^{\dagger}a_3$ commutes with the Hamiltonian,
thus we can diagonalize (\ref{7-3-7}) in the basis of eigenfunctions
of $B$. For any finite eigenvalue $b$ of $B$ the number of states is
finite, namely $(b+1)(b+2)/2$. Thus the infinite dimensional Hilbert space 
separates into an infinite set of finite dimensional subspaces, each 
subspace containing only vectors with a given eigenvalue $b$. These 
eigenfunctions are products of the number states $|n\rangle$ of each degree 
of freedom and can be characterized by a symbol $|n,m,l\rangle$ where we 
have $n$ bosons on site 1, $m$ bosons on site 2, and $l$ bosons 
on site 3. For a given value $b$ it follows that $l=b-m-n$.
So we can actually label each state by just two numbers $(n,m)$: 
$|n,m,(b-n-m)\rangle \equiv |n,m)$.
Note that the third site added to the dimer is different from the
first two sites. There is no boson-boson interaction on this site.
Thus site 3 serves simply as a boson reservoir for the dimer.
Dimer bosons may now fluctuate from the dimer to the reservoir.
The trimer has the same permutational symmetry as the dimer.

The matrix elements of (\ref{7-3-7}) between states from different $b$ 
subspaces vanish. Thus for any given $b$ the task amounts to 
diagonalizing a finite dimensional matrix. The matrix has a tridiagonal 
block structure, with each diagonal block being a dimer matrix 
(\ref{7-2-8}). The nonzero off-diagonal blocks contain interaction 
terms proportional to $\delta$. We consider symmetric $|\Psi\rangle_s$ and antisymmetric 
$|\Psi\rangle_a$ states. The structure of the corresponding symmetric and 
antisymmetric decompositions of $H$ is similar to $H$ itself.
\begin{figure}[!t]
\centering
\includegraphics[angle=-90,width=0.5\textwidth]{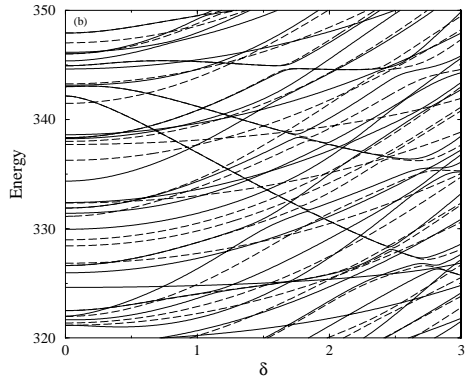}             
\caption{\label{fig7-4}A part of the eigenenergy spectrum 
of the quantum trimer as a function of $\delta$ 
with $b=40$ and $C=2$.
Lines connect data points for a given state. Solid lines - symmetric
eigenstates; thick dashed lines - antisymmetric eigenstates. 
Adapted from \cite{ff97jpc}.}
\end{figure}
In the following we will present results for $b=40$. We will also 
drop the first two terms of the RHS in (\ref{7-3-7}), because these only 
lead to a shift of the energy spectrum.  Since we evaluate the matrix 
elements explicitly, we need only a few seconds to obtain all eigenvalues 
and eigenvectors with the help of standard Fortran routines. In 
Fig.~\ref{fig7-4} we plot a part of the energy spectrum
as a function of $\delta$ for $C=2$ \cite{ff97jpc}.
As discussed above, the Hamiltonian decomposes into noninteracting 
blocks for $\delta=0$, each block corresponding to a dimer with a boson 
number between $0$ and $b$.  For $\delta \neq 0$ the block-block 
interaction leads to typical features in the spectrum, like, e.g., 
avoided crossings. 
The full quantum energy spectrum extends 
roughly over $10^3$, leading to an averaged spacing of order 
$10^0$. Also the upper third of the spectrum is diluted compared to the 
lower two thirds. 
The correspondence to the classical model is obtained with the use
of the transformation $E_{cl} = E_{qm}/b^2 + 1$ and for parameters
$C/b$ and $\delta/b$ (the classical value for $B$ is $B=1$).

\begin{figure}[!t]
\centering
\includegraphics[angle=-90,width=0.5\textwidth]{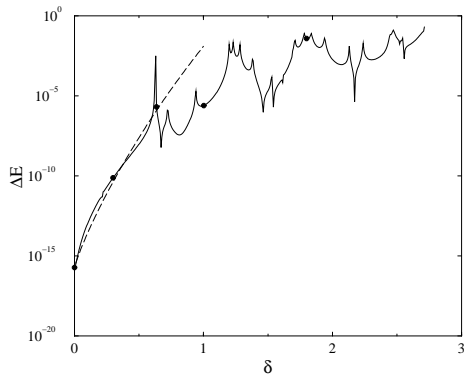} 
\caption{\label{fig7-5}Level splitting versus $\delta$ for a level
pair as described in the text. 
Solid line - numerical result.
Dashed line - semiclassical 
approximation. Filled circles - location of wave function
analysis in Fig.~\ref{fig7-6}. Adapted from \cite{ffo01prb}.}
\end{figure}

The main result of this computation so far is that tunneling pairs
of eigenstates of the dimer persist in the nonintegrable regime
$\delta \neq 0$. However at certain pair-dependent values of $\delta$ 
a pair breaks up. From the plot in Fig.~\ref{fig7-4} we cannot
judge how the pair splittings behave. In Fig.\ref{fig7-5} 
\index{Quantisation!Semi-Classical}
we
plot the pair splitting of the pair which has energy $\approx 342$ 
at $\delta=0$ \cite{ffo01prb}.
Denote with $x,y,z$ the
eigenvalues of the site number operators $n_1,n_2,n_3$. We may consider the
quantum states of the trimer at $\delta =0$ when $z$ is a good quantum number
and then follow the evolution of these states with increasing $\delta $. The
state for $\delta =0$ can be traced back to $C=0$ and be thus characterized in
addition by $x$ and $y$. 
The chosen pair states are then characterized by $x=26 (0)$, $y=0
(26)$ and $z=14$ for $C=\delta =0$. Note that this pair survives approximately
30 avoided crossings before it is finally destroyed at coupling strength
$\delta \approx 2.67$ as seen in Fig.~\ref{fig7-4}.

From Fig.~\ref{fig7-5} we find 
that the splitting rapidly increases gaining about eight
orders of magnitude when $\delta$ changes from 0 to slightly above 0.5. Then
this rapid but nevertheless smooth rise is interrupted by very sharp spikes
when the splitting $\Delta E$ rises by several orders of magnitude with
$\delta $ changing by mere percents and then abruptly changes in the opposite
direction sometimes even overshooting its pre-spike value. Such spikes, some
larger, some smaller, repeat with increasing $\delta $ until the splitting
value approaches the mean level spacing of order one. Only
then one may say that the pair is destroyed since it can be hardly
distinguished among the other trimer levels.

\begin{figure}[!t]
\centering            
\includegraphics[angle=-0,width=0.3\textwidth]{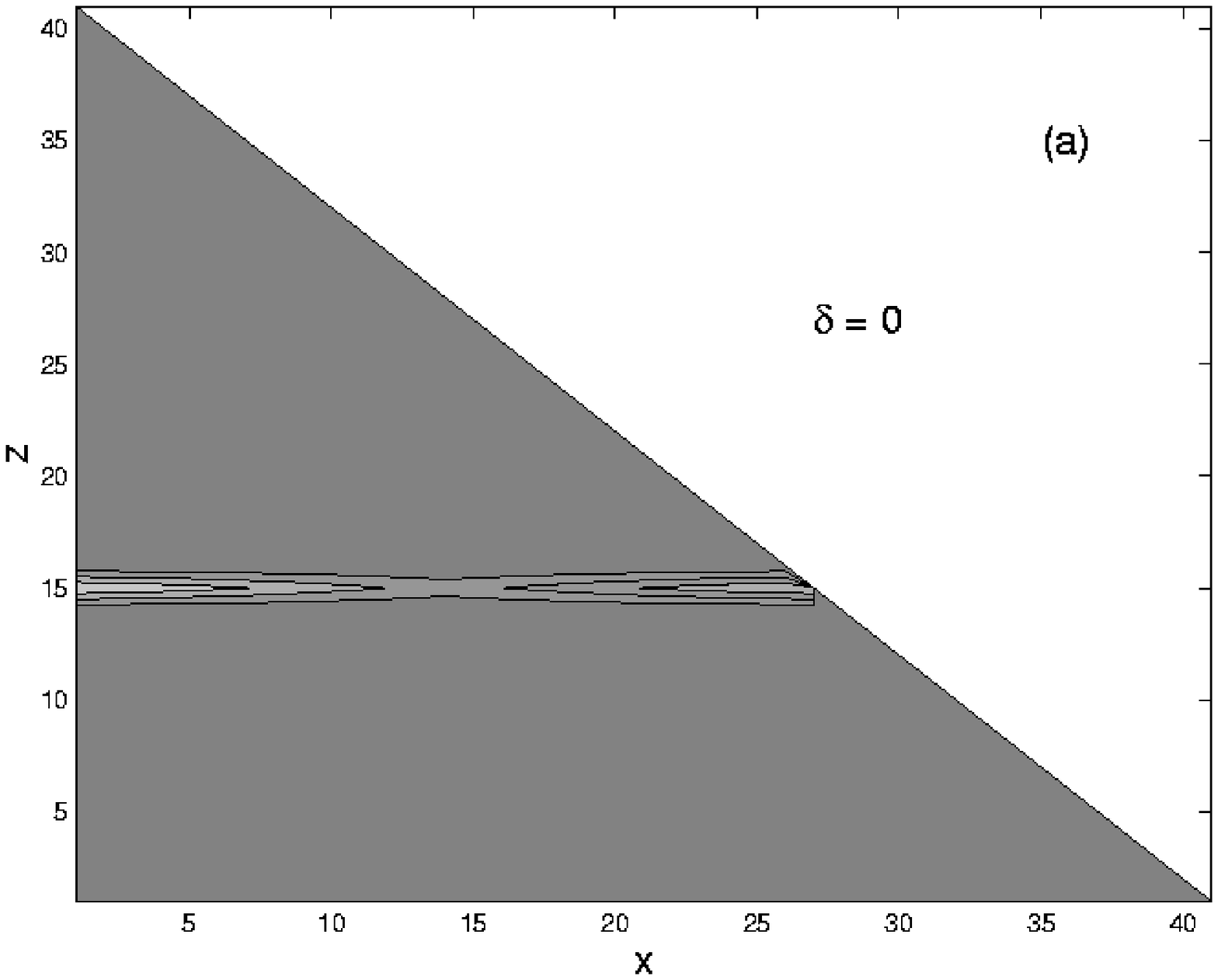}
\includegraphics[angle=-0,width=0.3\textwidth]{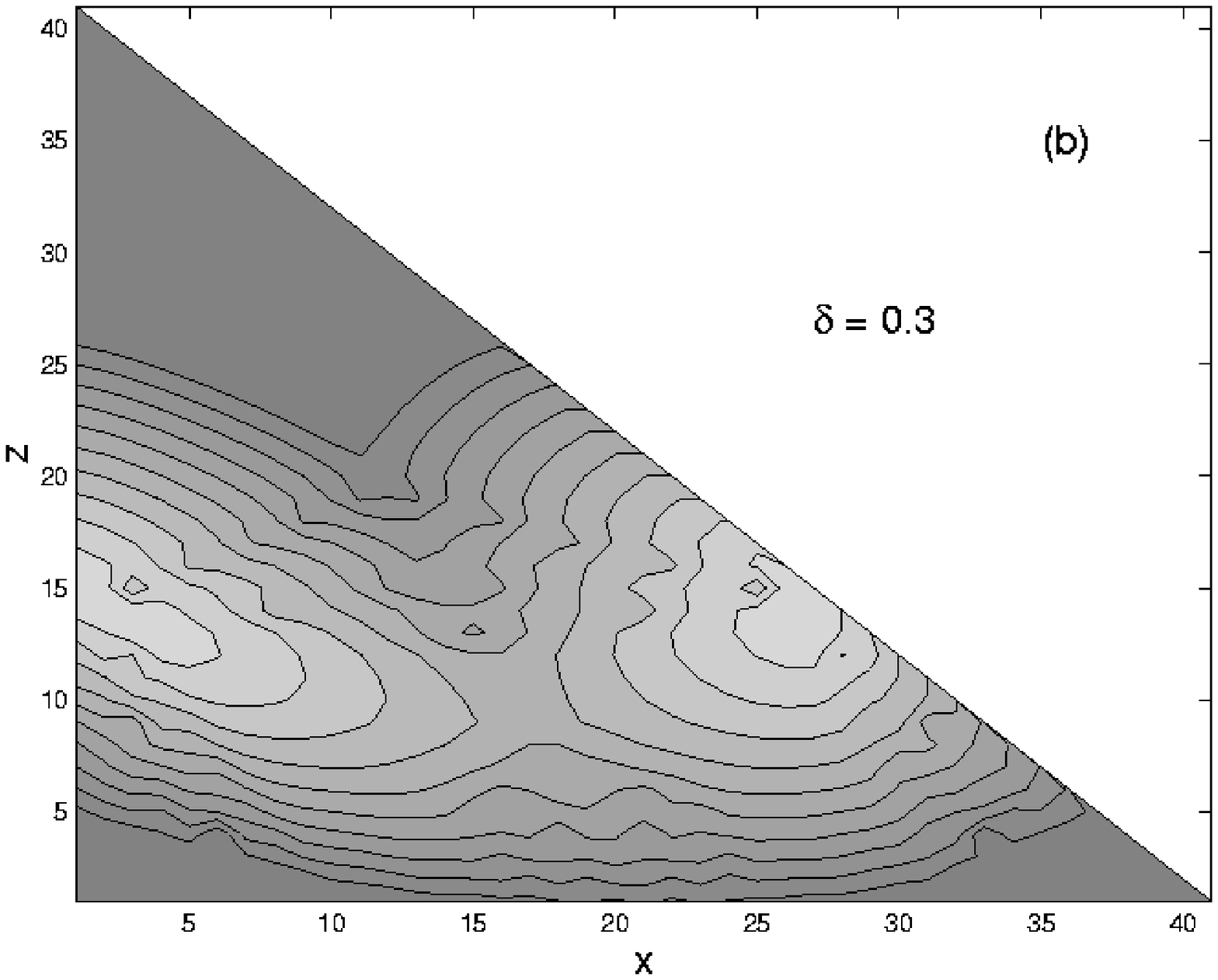}
\includegraphics[angle=-0,width=0.3\textwidth]{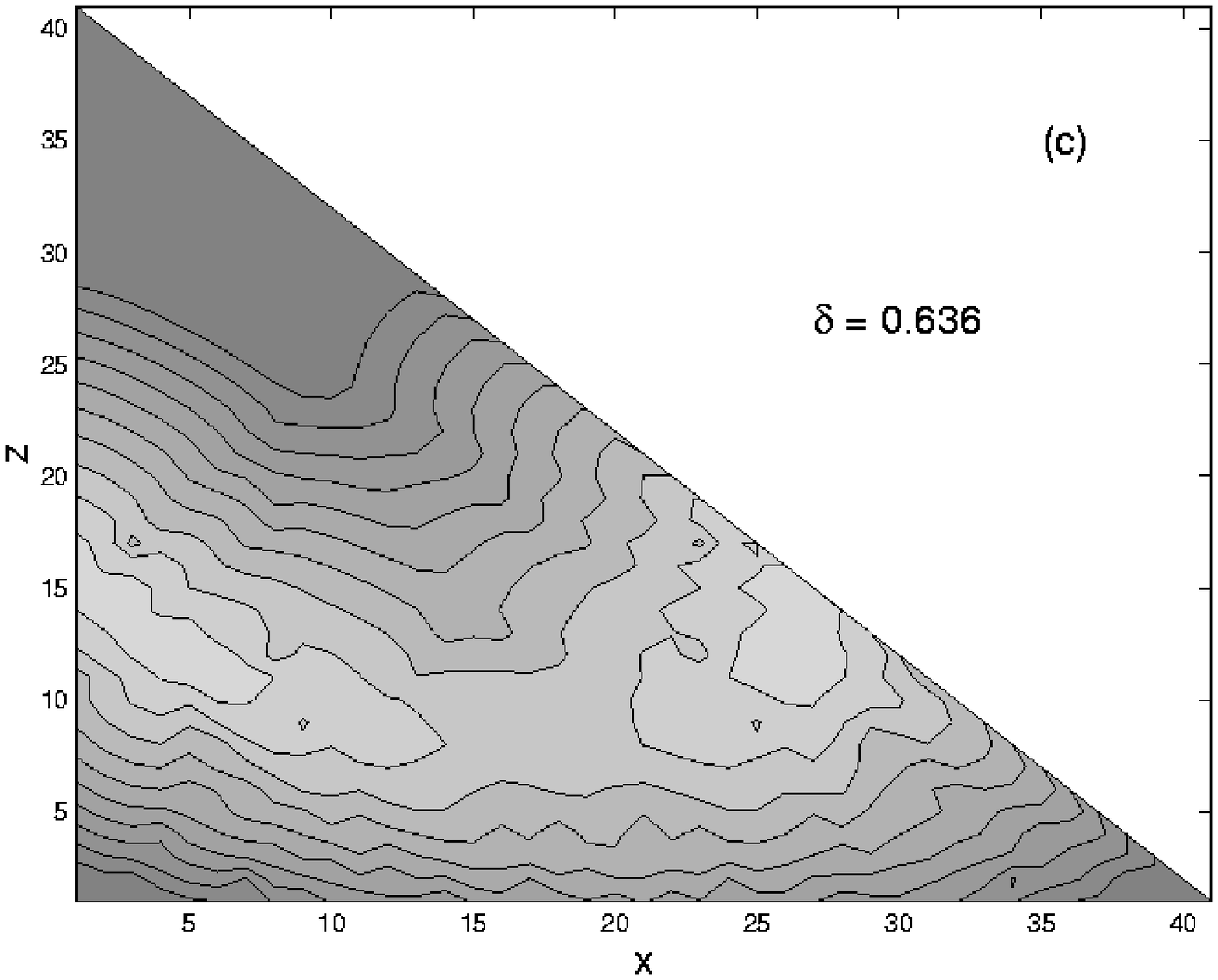}
\\
\includegraphics[angle=-0,width=0.3\textwidth]{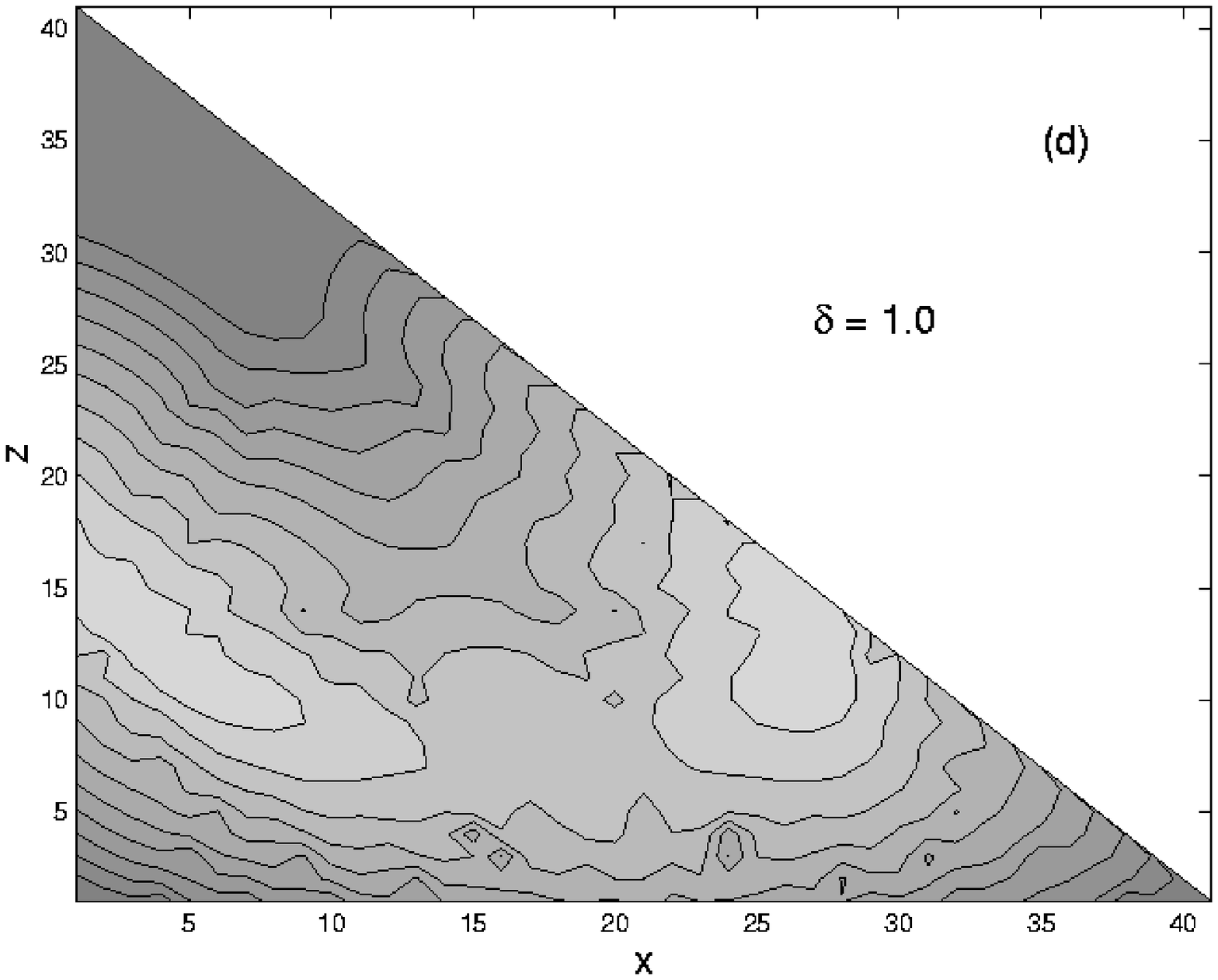}
\includegraphics[angle=-0,width=0.3\textwidth]{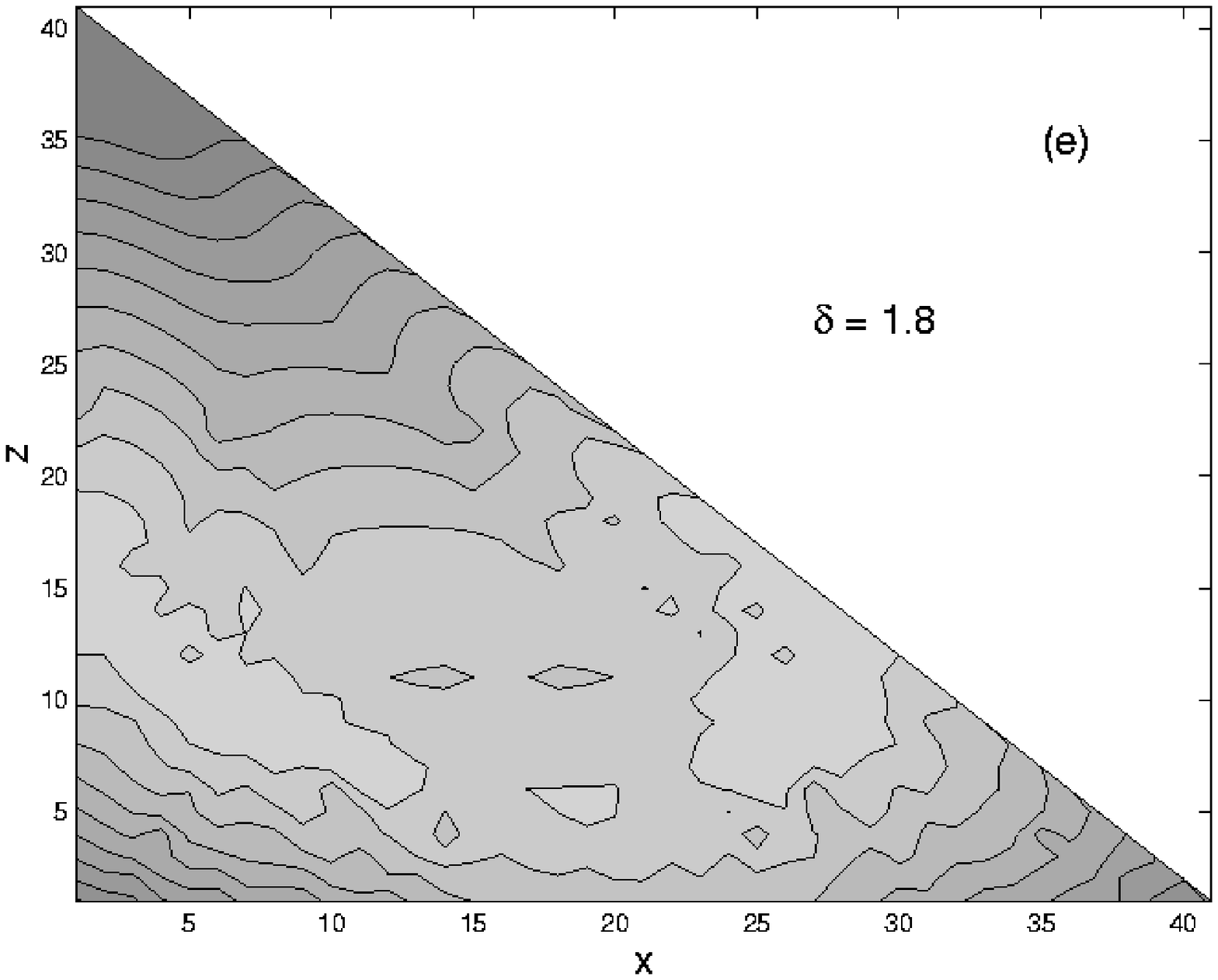}
\caption{\label{fig7-6}Contour plot of the logarithm of 
the symmetric eigenstate of the chosen
tunneling pair (cf. Fig.~\ref{fig7-4}) for five different values of
$\delta=0,\;0.3,\;0.636,\;1.0,\;1.8$ (their location is indicated by
filled circles in Fig.~\ref{fig7-5}). 
(a): three equidistant grid lines are used; (b-e): ten grid
lines are used. Minimum value of squared wave function is $10^{-30}$, maximum
value is about 1.  
Adapted from \cite{ffo01prb}.}
\end{figure}

\begin{figure}[!t]
\centering          
\includegraphics[angle=-90,width=0.5\textwidth]{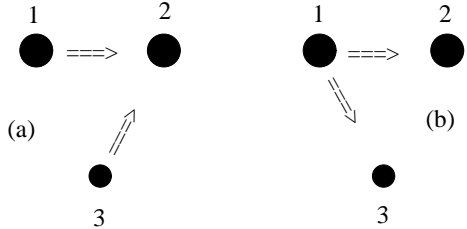}
\caption{\label{fig7-7}Order of tunneling in the trimer. Filled large circles - sites 1 and 2, filled
small circle - site 3. Arrows indicate direction of transfer of particles.
Adapted from \cite{ffo01prb}.}
\end{figure}

Another observation is presented in Fig.~\ref{fig7-6} \cite{ffo01prb}. We plot the intensity
distribution of the logarithm of the squared symmetric wave function of our
chosen pair for five different values of $\delta=0\;,\;0.3\;,\;0.636\;,\;
1.0\;,\;1.8$ (their locations are indicated by filled circles in 
Fig.~\ref{fig7-5}). We
use the eigenstates of $B$ as basis states. They can be represented
as $|x,y,z>$ where $x,y,z$ are the particle numbers on sites 1, 2, 3,
respectively. Due to the commutation of $B$ with $H$ two site occupation
numbers are enough if the total particle number is fixed. Thus the final
encoding of states (for a given value of $b$) can be chosen as $|x,z)$. The abscissa in Fig.~\ref{fig7-6} 
is $x$ and the
ordinate is $z$. Thus the intensity plots provide us with information about
the order of particle flow in the course the tunneling process.
For $\delta =0$ (Fig.~\ref{fig7-6}(a)) the only possibility for the 26 particles on site 1
is to directly tunnel 
to site 2. Site 3 is decoupled with its 14 particles not
participating in the process. The squared wave function takes the form of a
compact rim in the $(x,z)$ plane which is parallel to the $x$ axis. Nonzero
values of the wave function are observed only on the rim. This direct tunneling
has been described in chapter \ref{sec7.2}. When switching on some
nonzero coupling to the third site, the particle number on the dimer (sites
1,2) is not conserved anymore. The third site serves as a particle reservoir
which is able either to collect particles from or supply particles to the
dimer. This coupling will allow for nonzero values of the wave function away
from the rim. But most importantly, it will change the shape of the rim. We
observe that the rim is bended down to smaller $z$ values with increasing
$\delta$. That implies that the order of tunneling (when, e.g., going from
large to small $x$ values) is as follows: first, some particles tunnel from
site 1 to site 2 and simultaneously from site 3 to site 2
(Fig.~\ref{fig7-7}(a)). 
Afterwards particles flow from site 1 to both sites 2 and 3
(Fig.~\ref{fig7-7}(b)). With increasing $\delta$ the structure of the wave function
intensity becomes more and more complex, possibly revealing information about
the classical phase space flow structure.
Thus we observe three intriguing features. First, the tunneling splitting
increases by eight orders of magnitude when $\delta$ increases from zero to
0.5. 
This seems to be unexpected, since at those values perturbation theory in
$\delta$ should be applicable (at least Fig.~\ref{fig7-4} 
indicates 
that this should be true for the levels themselves). 
The semiclassical explanation of this result was obtained 
in \cite{ffo01prb}.
\\
The second observation is
that the tunneling begins with a flow of particles from the bath (site 3)
directly to the empty site which is to be filled (with simultaneous flow from
the filled dimer site to the empty one). At the end of the tunneling process
the initially filled dimer site is giving particles back to the bath
site. Again this is an unexpected result, since it implies that the particle
number on the dimer is increasing during the tunneling, which seems to
decrease the tunneling probability, according to the results for an isolated
dimer. These first two results are closely  connected (see 
\cite{ffo01prb}
for a detailed explanation).
\begin{figure}[!t]
\centering           
\includegraphics[angle=-90,width=0.3\textwidth]{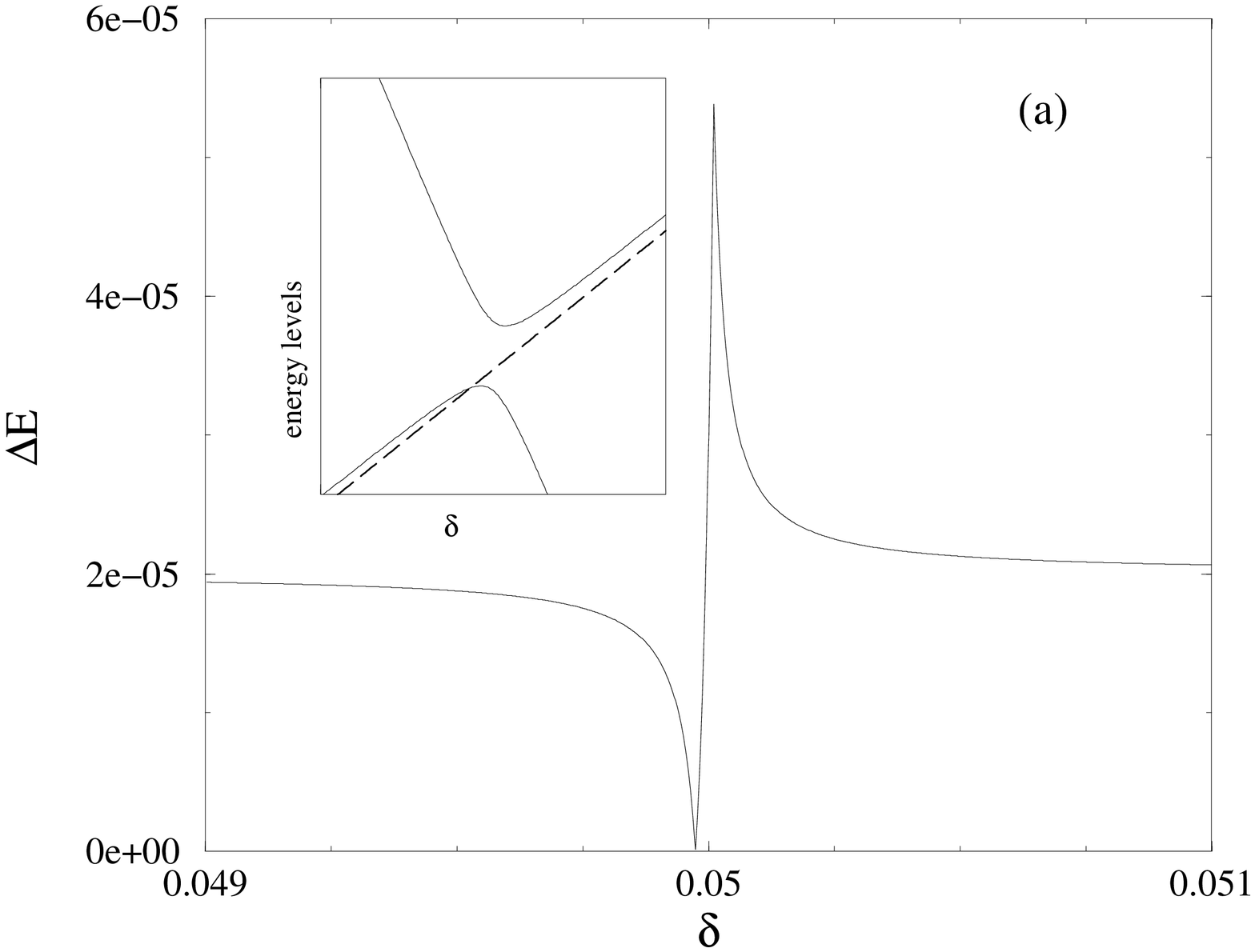}
\includegraphics[angle=-90,width=0.3\textwidth]{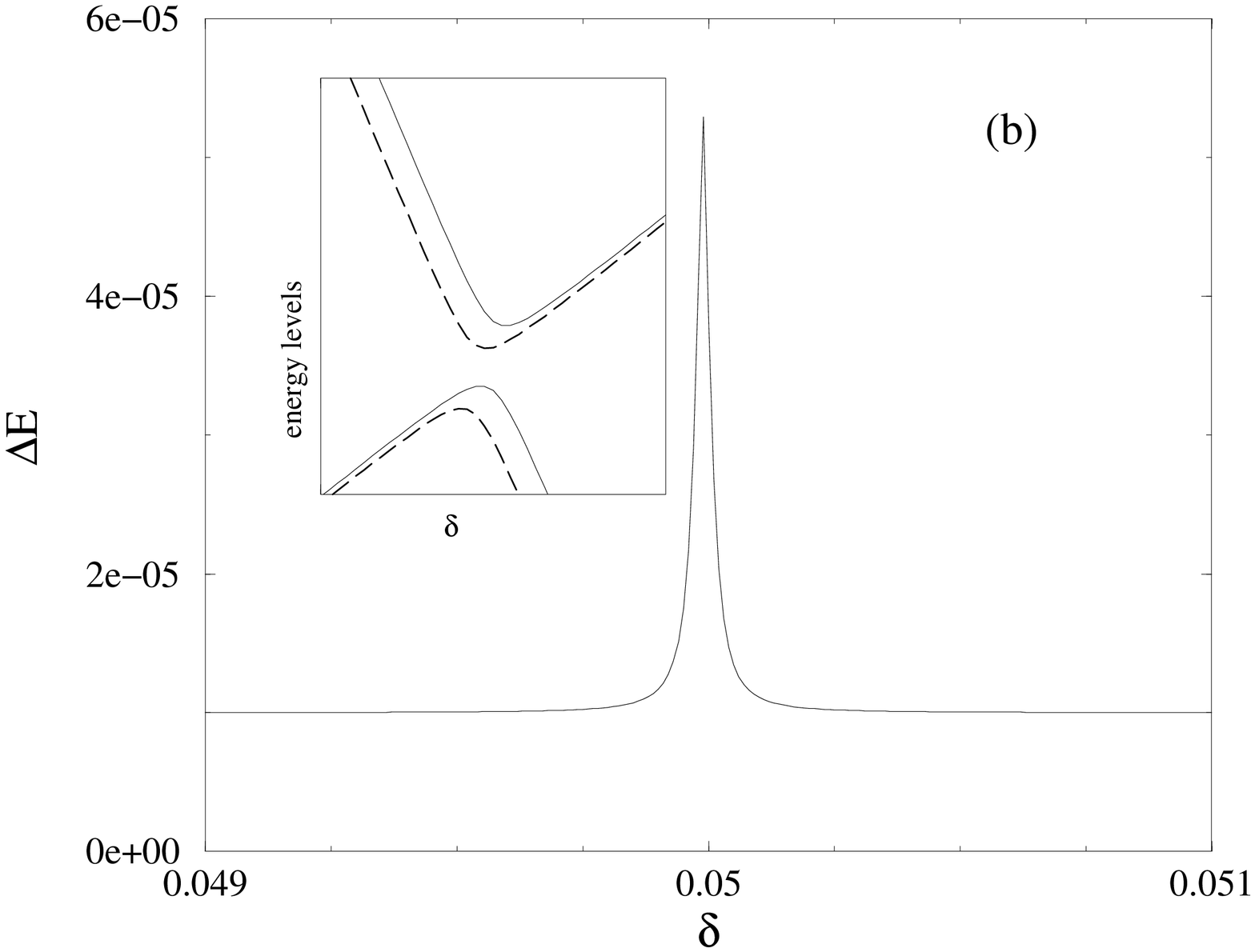}
\\
\includegraphics[angle=-90,width=0.3\textwidth]{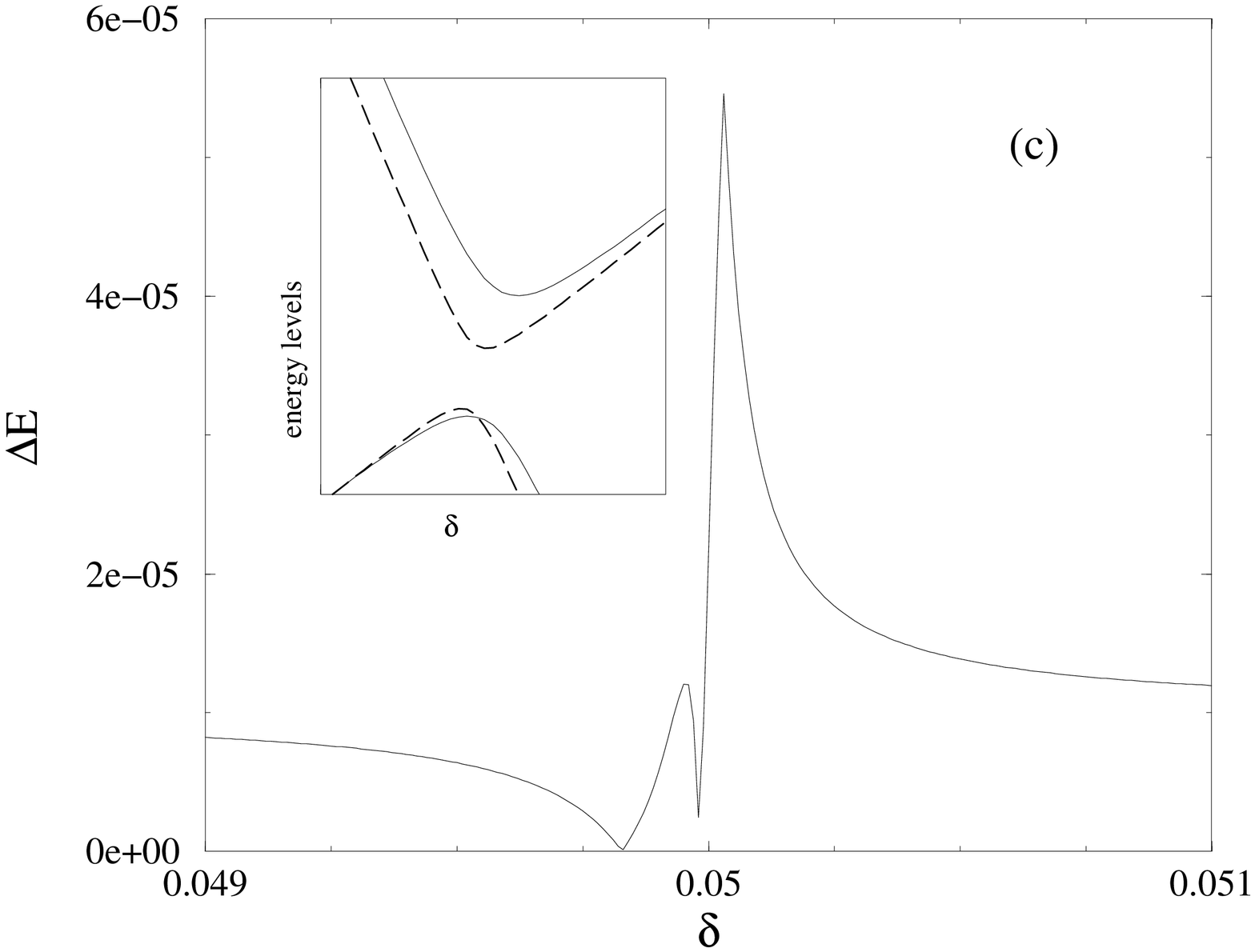}
\includegraphics[angle=-90,width=0.3\textwidth]{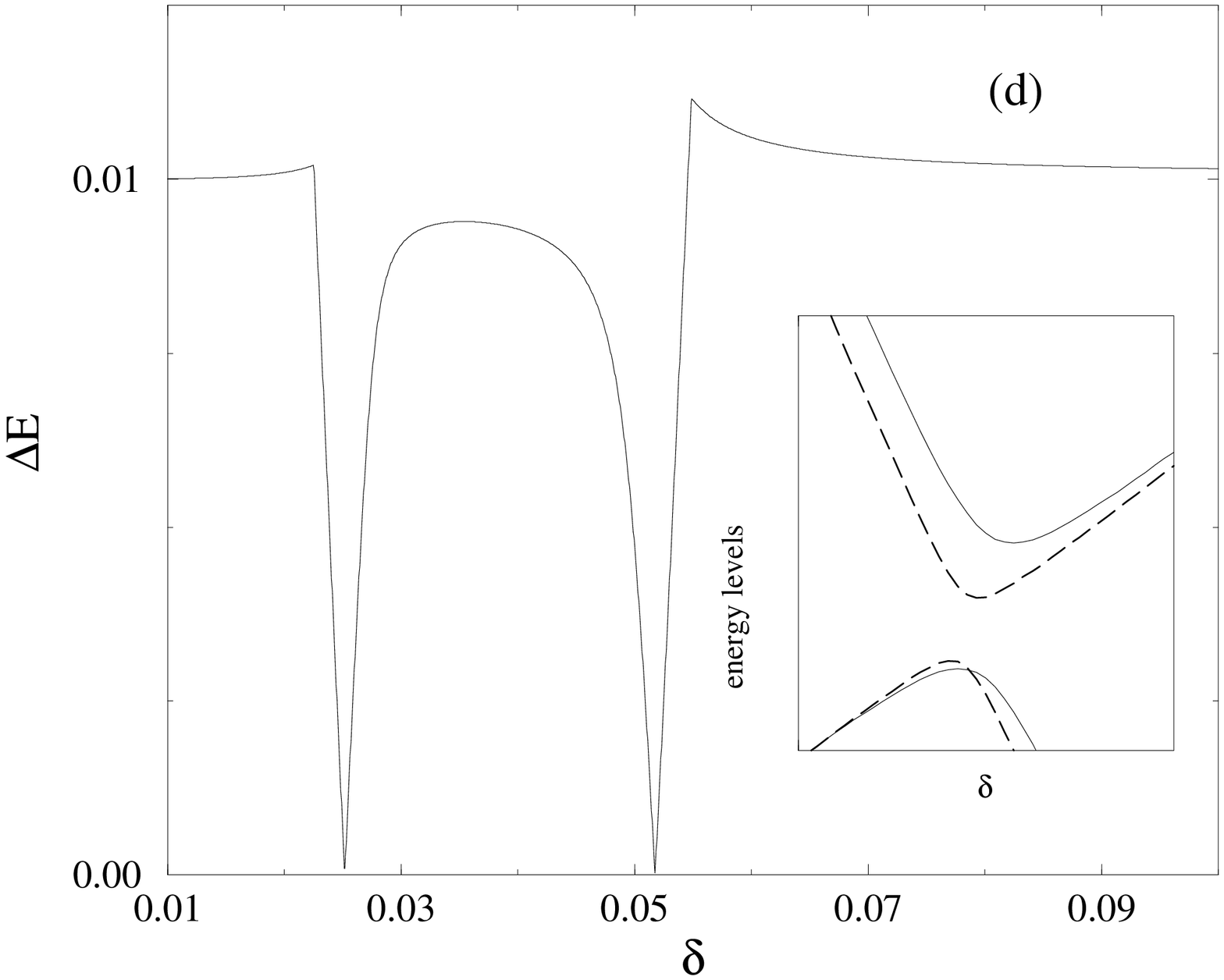}
\caption{\label{fig7-8}Level splitting variation at avoided crossings. 
Inset: Variation of individual eigenvalues participating in the avoided
crossing. Solid lines - symmetric eigenstates, dashed lines - antisymmetric
eigenstates. 
Adapted from \cite{ffo01prb}.}
\end{figure}
\\
The third result concerns the resonant structure on top of the smooth
variation in Fig.~\ref{fig7-5}. 
The resonant enhancements and suppressions of tunneling
are related to avoided crossings. Their presence implies that a fine tuning of
the system parameters may strongly suppress or enhance tunneling which may be
useful for spectroscopic devices.
In Fig.~\ref{fig7-8} we show the four various possibilities of
avoided crossings between a pair and a single level and between
two pairs, and the schematic outcome for the tunneling splitting \cite{ffo01prb}.
If the interaction to further more distant states in the spectrum is added,
the tunneling splitting can become exactly zero \cite{rapsf06pra} for some
specific value of the control parameter. In such a rare situation the tunneling
is suppressed for all times.

\subsection{Quantum roto-breathers}
\label{sec7.4}

When discussing classical breather solutions we have been touching
some aspects of roto-breathers, including their property of being
not invariant under time reversal symmetry. In a recent study
Dorignac et al  have performed \cite{jdsf02prb} an analysis of the
corresponding quantum roto-breather properties in a dimer with
the Hamiltonian
\begin{equation}
H = \sum_{i=1}^{2} \left\{ \frac{p_i^2}{2}+\alpha (1-\cos x_i) \right\}
+\varepsilon (1-\cos (x_1-x_2))
\;.
\label{7-8-1}
\end{equation}
The classical roto-breather solution consists of one pendulum 
rotating and the other oscillating with a given period $T_b$.
Since the model has two symmetries - permutation of the 
indices and time-reversal symmetry - which may be both
broken by classical trajectories, the irreducible 
representations of quantum eigenstates contain four symmetry
sectors (with possible combinations of
symmetric or antisymmetric states with respect to the two
symmetry operations). 
Consequently, a quantum roto-breather state is belonging
to a quadruplet of weakly split states rather than to a 
pair as discussed above.
\begin{figure}[!t]
\centering
\includegraphics[angle=-0,width=0.6\textwidth]{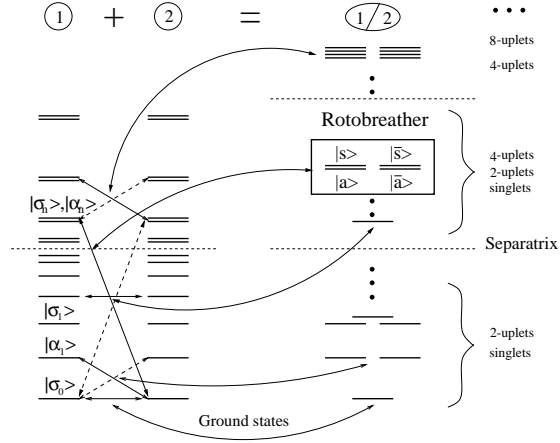}            
\caption{\label{fig7-9}Schematic representation of the sum of two pendula spectra. Straight solid arrows indicate the levels to be added and dashed arrows the symmetric (permutation) operation. The result is indicated in the global spectrum by a curved arrow. The construction of the quantum roto-breather state is explicitly represented. Adapted from \cite{jdsf02prb}.}
\end{figure}
The schematic representation of the appearance of such a quadruplet
is shown in Fig.~\ref{fig7-9} \cite{jdsf02prb}.
The obtained quadruplet has an
additional fine structure as compared to the tunneling pair
of the above considered dimer and trimer. 
The four levels in the quadruplet define three characteristic
tunneling processes. Two of them are energy or momentum transfer
from one pendulum to the other one, while the third one
corresponds to total momentum reversal (which restores 
time reversal symmetry).
\begin{figure}[!t]
\centering         
\includegraphics[angle=-0,width=0.5\textwidth]{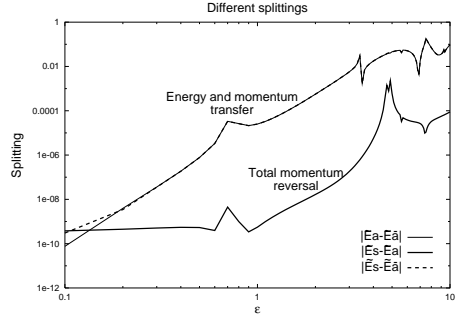} 
\caption{\label{fig7-10}Dependence of different splittings of a
quadruplet on $\varepsilon$. Only three of them have been displayed, each being associated with a given tunneling process. Adapted from \cite{jdsf02prb}.}
\end{figure}
The dependence of the corresponding tunneling rates on the
coupling $\varepsilon$ is shown for a specific quadruplet from
\cite{jdsf02prb} in Fig.~\ref{fig7-10}.
For very weak coupling $\varepsilon \ll 1$ the fastest tunneling
process will be momentum reversal, since tunneling between 
the pendula is blocked. However as soon as the coupling is increased,
the momentum reversal turns into the slowest process, with
breather tunneling from one pendulum to the other one being
orders of magnitude faster. Note that again resonant features
on these splitting curves are observed, which are related to
avoided crossings.

\subsection{Large lattices with fluctuating numbers of quanta}

\begin{figure}[!t]
\centering          
\includegraphics[angle=-0,width=0.5\textwidth]{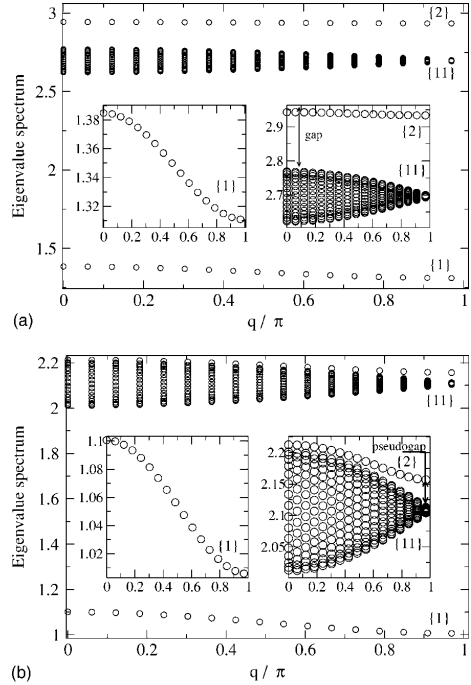} 
\caption{\label{fig7-12}Eigenspectrum of a chain with 33 sites for parameters (a) $C=0.05$, $v_4=0.2$, and (b) $C=0.05$, $v_4=0.02$. The inserts show magnifications of the fundamental
branch (left) and overtone region (right). The quantum breather branch is marked by
(2), and the two-phonon band by (11).
Adapted from \cite{lp05prb1}.}
\end{figure}

\begin{figure}[!t]
\centering           
\includegraphics[angle=-0,width=0.99\textwidth]{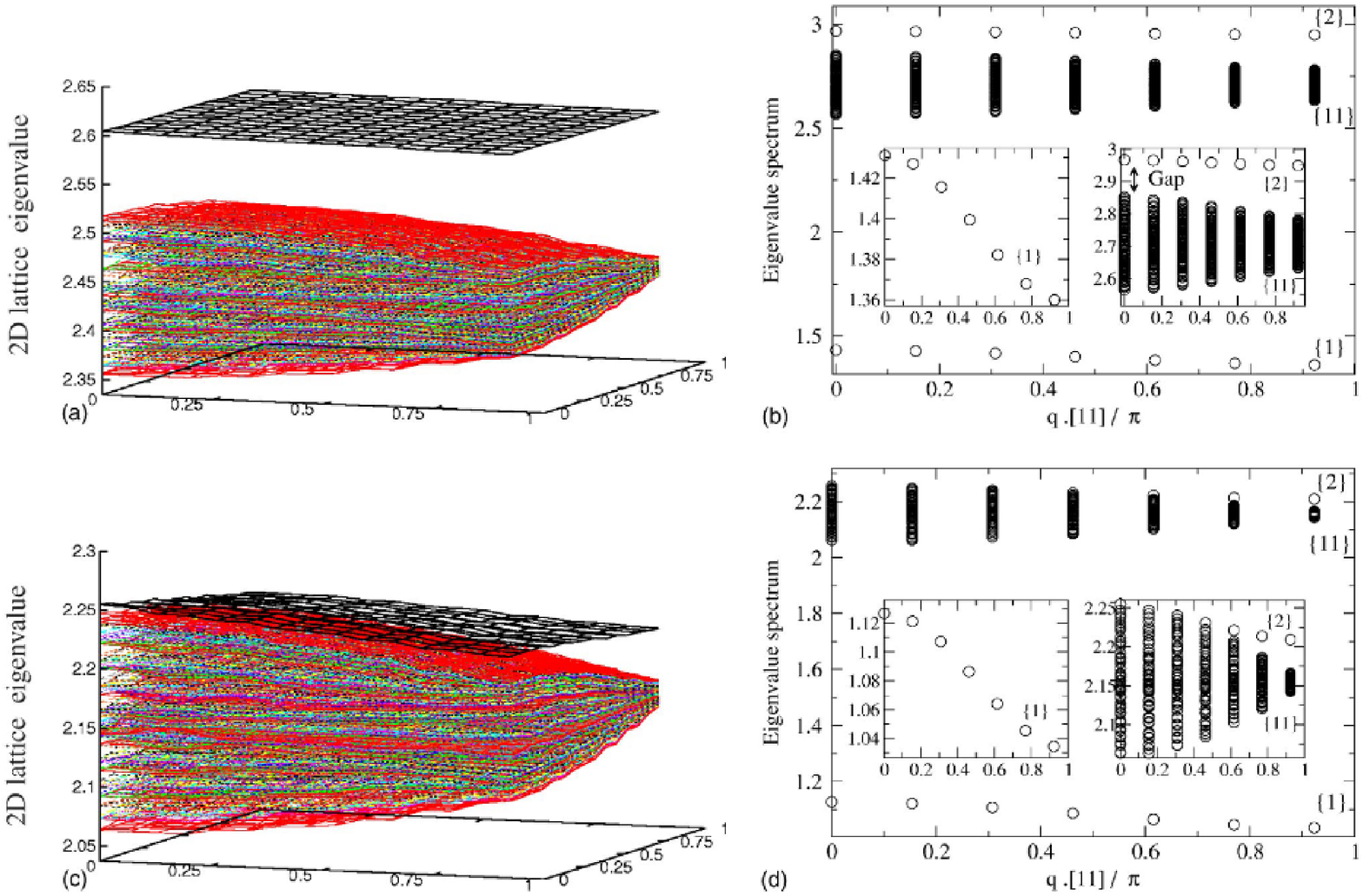} 
\caption{\label{fig7-13}Eigenspectrum of a lattice with $13\times 13$ sites for parameters (top) $C=0.025$, $v_4=0.1$, and
(bottom) $C=0.025$, $v_4=0.025$. 
Left plots - spectra over the whole Brilloin zone.
Right plots - profiles of the spectra along the direction [11]. The insets
show the magnifications of the phonon branch (left) and the quantum breather
energy region (right).
Adapted from \cite{lp05prb1}.}
\end{figure}

A number of publications are devoted to the properties of quantum
breathers in chains and two-dimensional lattices of coupled anharmonic
oscillators. For the respective one-dimensional case, the 
Hamiltonian is given by
\begin{equation}
H = \sum_n \left[ \frac{1}{2}p_n^2 + V(x_n) + W(x_n - x_{n-1}) \right]\;.
\label{7.8-2}
\end{equation}
Here $V(x)=\frac{1}{2}x^2 + \frac{1}{4}v_4 x^4$ (or similar) and the nearest neighbour
coupling $W(x)=\frac{1}{2}C x^2$.
The classical version of such models conserves only the energy, but not any equivalent
of a norm. Therefore, no matter whether one uses creation and annihilation operators
of the harmonic oscillator \cite{lp05prb1}, or similar operators which diagonalize the 
single anharmonic oscillator problem \cite{lp05prb2}, the resulting Hamiltonian
matrix will not commute with the corresponding number operator. 
Calculations will typically be restricted to 4-6 quanta, and lattice sizes
of the order of 30 for $d=1$, $13\times 13$ for $d=2$ \cite{lp05prb1}.
With these parameters one can calculate properties of quantum breather states, which correspond
to typically two quanta which are bound (with unavoidable states with different
number of quanta, contributing as well). 
For large enough $v_4$ a complete gap opens between the two-quanta continuum
and quantum breather states \cite{wgbs96,lp05prb1} (Figs.\ref{fig7-12} and \ref{fig7-13}).
When decreasing 
the anharmonic constant $v_4$, Proville found, that the gap closes for certain wave numbers,
but persists for others, becoming a pseudogap \cite{lp05epl,lp05prb1} (Fig.\ref{fig7-12}).

\begin{figure}[!t]
\centering            
\includegraphics[angle=-0,width=0.7\textwidth]{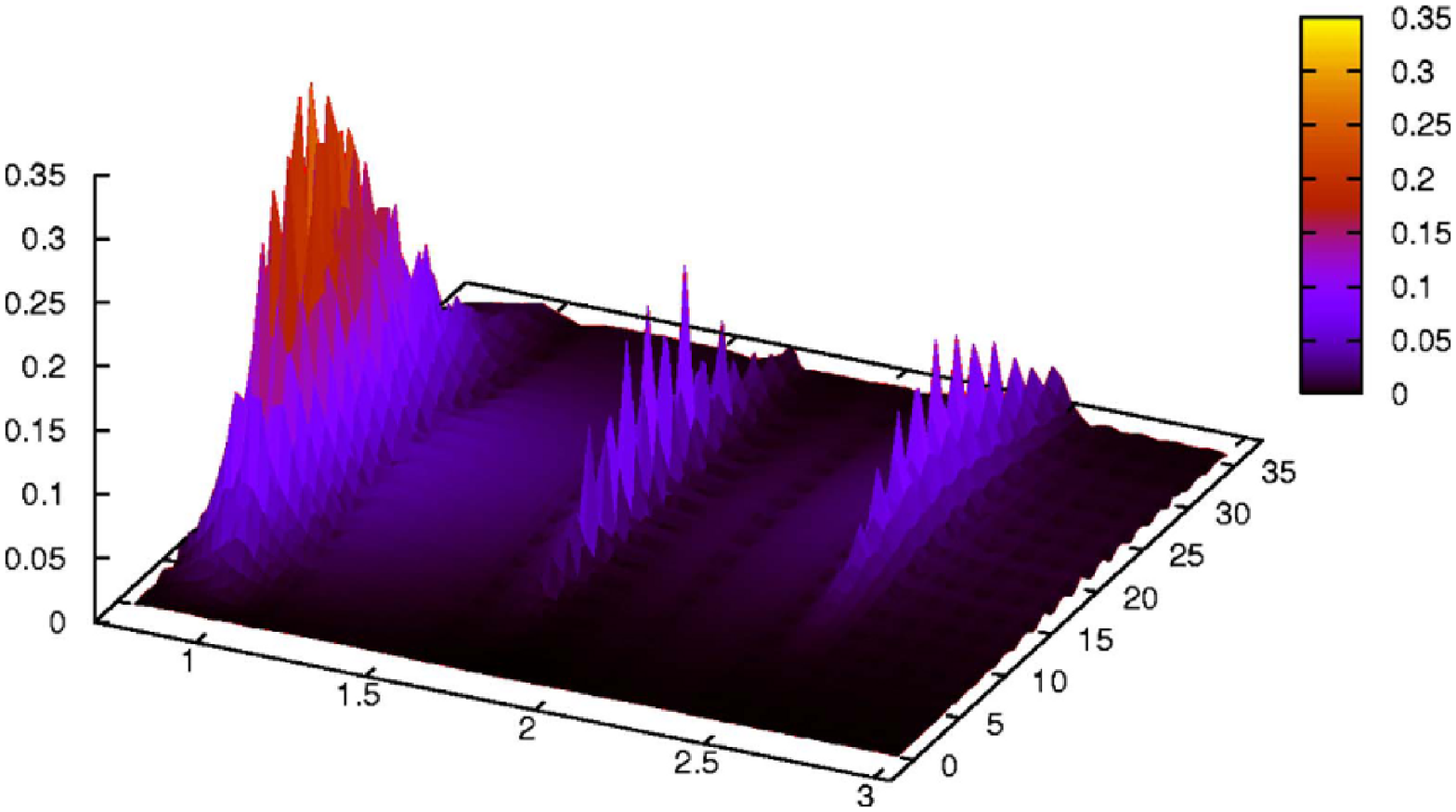} 
\caption{\label{fig7-14}A 3D plot of the inelastic structure factor $S(q,\omega)$ as a function of the dimensionless energy transfer $ 0 \leq \omega \leq 3$ and the scalar product of the transfer momentum $\bf{q}$ and the polarization $\bf{u}$.  
Adapted from \cite{lp05prb2}.}
\end{figure}
Involved calculations of the dynamical structure factor (e.g. available by neutron
scattering in crystals) have shown, that signatures
of quantum breathers are imprinted in these integral characteristics of 
the underlying lattice dynamics \cite{wgbs96,lp05prb2}, yet the working out of these
differences may become a subtle task (see Fig.\ref{fig7-14} for an example). 

Finally, Fleurov et al \cite{vfrssf98} estimated the influence of the tunnel splitting 
of a dimer, when embedded in an infinite chain. This situation is close to the tunneling
of a very localized DB, so that the nonlinearity (interaction between bosons) can be taken
into account only on the two sites, which participate in the tunneling, while the nonlinearity
can be neglected on all other sites. Using path integral techniques, the computed tunneling
splitting has been shown to become smaller as compared to the case of an isolated dimer. 
This is due to the fact, that a DB in an infinite chain has a core and a localized tail.
That tail has to be carried through the tunneling process as well, and in analogy with
a single particle tunneling in a double well, the tail increases the effective mass of such a particle.
Consequently the exponential tail of a DB in an infinite chain tends to decrease its ability
to perform quantum tunneling motion, yet it never leads to a full suppression of tunneling \cite{vfrssf98}.

\section{Quantum breather properties}

\subsection{Evolution of quantum localized states}
\label{sec7.5}

\begin{figure}[!t]
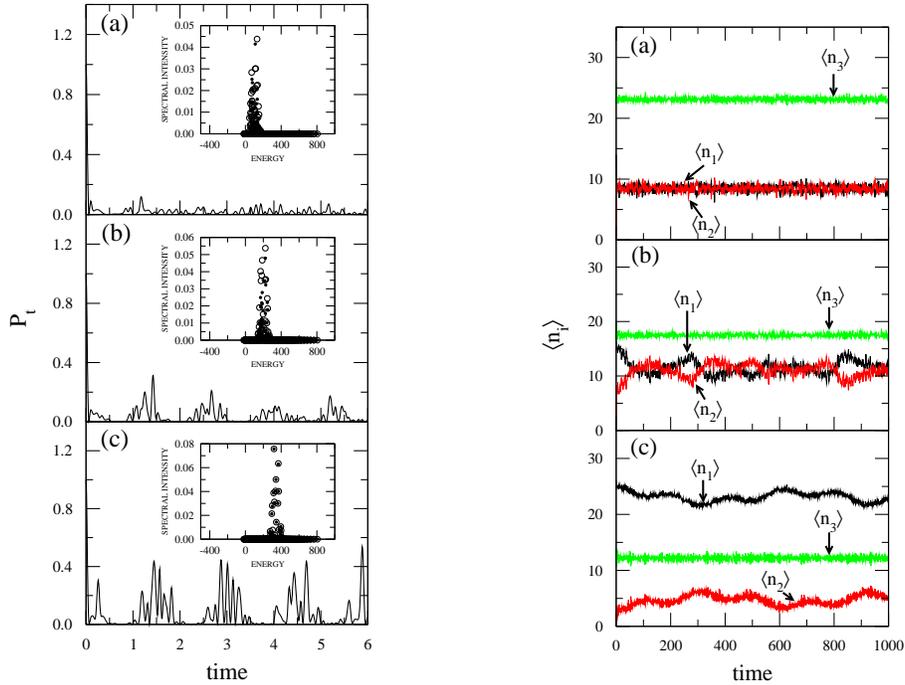
 
\includegraphics[angle=-0,width=0.4\textwidth]{fig7-11a.eps}
\hspace*{2cm} 
\includegraphics[angle=-0,width=0.4\textwidth]{fig7-11b.eps} 
\caption{\label{fig7-11}Left plot: Survival probability of the initial state $|\Psi_0 \rangle = |20+\nu,0,20-\nu\rangle$.
$\nu=$ (a) -6, (b) 0, and (c) 6. Insets: spectral intensity of the initial
state $|\Psi_0 \rangle$. Filled circles - symmetric eigenstates; open circles -
antisymmetric eigenstates. Right plot: Time evolution of expectation values of the number of bosons at each site of the trimer for different initial states 
$|\Psi_0 \rangle = |20+\nu,0,20-\nu\rangle$.
$\nu=$ (a) -6, (b) 0, and (c) 6. Adapted from \cite{rapsf06pra}.}
\end{figure}

Suppose that we initially excite only one site in the trimer from above.
If this initial state has strong enough overlap with tunneling pair eigenstates,
its evolution in time should show distinct properties as compared to
the case when the overlap is vanishing, or when there are simply no
tunneling pair states available.
Several results have been reported. First, a quantum echo was observed in \cite{ff97jpc}
by calculating the survival probability of the initial state as a function of time.
That quantity measures the probability to find the system in the initial state at
later times. If the initial state has strong overlap with many eigenstates, 
it is expected to quickly decohere into these different states. Yet, if 
a substantial overlap with quantum breathers takes place, the survival probability
first rapidly decays to zero, but echoes up after regular time intervals
(Fig.\ref{fig7-11}, left plot).
If one simply measures the dependence of the number of quanta, then a similar
situation will show up with a very slow beating of the occupation numbers
in time, if the overlap of the initial state and a tunneling pair is strong \cite{rapsf06pra,gkarbvmk03}
(see Fig.\ref{fig7-11}, right plot).

Suppose we have a large lattice, and put initially many quanta on one site.
Then any tunneling of this packet as a whole will occur on very long time scales.
On time scales much shorter, we may describe the excitation as a classical
discrete breather state plus a small perturbation. Treating that perturbation
quantum mechanically, one could expect that the time-periodic DB acts as a constant
source of quantum radiation for the quantized phonon field. It turns out to be impossible,
for very much the same reasons as in the purely classical treatment (see \cite{sfvfag05prb}).
This result implies, that there is almost no other source of decay for a localized initial
state in a quantum lattice, but to slowly tunnel as a whole along the lattice,
if nonlinearities allow for the formation of exact classical DB states \cite{Schulman2006PRL96}.
Numerical calculations for such a case, but with few quanta, were performed by Proville \cite{lp06pd},
and, similar to the above trimer discussion, showed that if quantum breather states
exist in the system, then localized excitations stay localized for times which are much longer
than the typical phonon diffusion times in the absence of anharmonicity.

\subsection{Splitting and correlations}

QBs are nearly degenerate eigenstates. For the dimer and the trimer, they come in symmetric-antisymmetric pairs. So one may compute the nearest neighbor energy spacing (tunneling splitting) between pairs of symmetric-antisymmetric eigenstates in order to identify QBs. Since QBs correspond to classical orbits that are characterized by energy localization, they may be identified by defining correlation functions. For large lattices it has been shown that QBs have exponentially localized correlation functions, in full analogy to their classical counterparts.

For the dimer and the trimer, the correlation functions may be defined as follows:
\begin{equation}\label{eq:f12dimer}
f_{\mu}(1,2) = \langle\hat{n}_1\hat{n}_2\rangle_{\mu}, \;\;\;\;\;\;\;
f_{\mu}(1,1) = \langle\hat{n}_1^2\rangle_{\mu},
\end{equation}
where $\hat{n}_i=\hat{a}_i^{\dagger}\hat{a}_i$, and $\langle\hat{A}\rangle_{\mu} =
\langle\chi_{\mu}|\hat{A}|\chi_{\mu}\rangle$,
$\{|\chi_{\mu}\rangle\}$ being the set of eigenstates of the system.
The ratio $0 \leq f_{\mu}(1,2)/f_{\mu}(1,1) \leq 1$ measures the site correlation of quanta:
it is small when
quanta are site-correlated (i.e. when 
many quanta are located on one site there are almost none on the other one) 
and close to unity otherwise.
\begin{figure}[!t]
\begin{center}
\includegraphics[width=3.in]{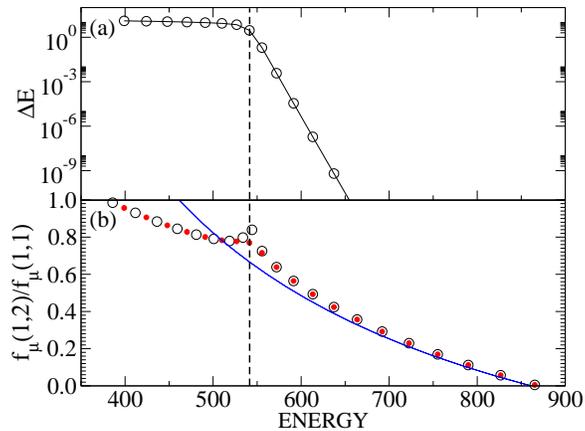}
\caption{\label{corrdimer}(a) Energy splitting and (b) correlation function
vs. energy of the eigenstates of the dimer (open circles, symmetric
eigenstates; solid circles, antisymmetric eigenstates). The vertical dashed
line marks the energy threshold for appearance of QB states. The thin solid line in (a) is a guide for the eye, whereas in (b) it is the estimation using eq. (\ref{eq:corrdimertheo}). Here $b=40$ and $C=2$.}
\end{center}
\end{figure} 

For the dimer case, the relation $b=n_1+n_2$, leads to
\begin{equation}\label{f12dimer2}
f_{\mu}(1,2) = b\langle\hat{n}_1\rangle_{\mu} -  \langle\hat{n}_1^2\rangle_{\mu}.
\end{equation}
In Fig. \ref{corrdimer}-left we show the  energy splitting and the correlation function of
the eigenstates. We see that beyond a threshold (dashed line), the splitting drops exponentially fast with energy. The corresponding pairs of eigenstates, which are tunneling pairs,
are site correlated. Thus they are QBs. Their correlation functions
show a fast decrease for energies above the threshold. In these states
many quanta are localized on one site of the dimer and the tunneling time of
such an excitation from one site to the other (given by the inverse energy
splitting between the eigenstates of the pair) is exponentially large. As
shown in Ref. \cite{afko96prl}, this energy threshold is close to the
threshold for the existence of DBs in the corresponding classical model. 

QBs are close to symmetric (S) and antisymmetric (A) eigenstates of the $C=0$ case given by
\begin{equation}\label{eq:dimerbasis}
|n_1,n_2\rangle_{S,A}=\frac{1}{\sqrt{2}}(|n_1,n_2\rangle \pm |n_2,n_1\rangle),
\end{equation}
with $n_{1,2}\gg n_{2,1}$.
So we may estimate the dependence of the correlation functions on $n_1$ using the eigenstates (\ref{eq:dimerbasis}) and $b=n_1+n_2$. The result is:
\begin{equation}\label{eq:corrdimertheo}
\frac{f(1,2)_{n_1}}{f(1,1)_{n_1}} = \frac{2n_1(b-n_1)}{n_1^2+(b-n_1)^2},
\end{equation}
where we note that it is equal to unity when $n_1=b/2$, and vanishes when $n_1=0,b$.

Using the relation between the eigenenergy $\varepsilon$ of the $C=0$ case and the number $n_1$ ($=1,2,\ldots ,b$)
\begin{equation}
\varepsilon_{n_1} = \frac{5}{4} + \frac{3}{2}b + \frac{1}{2}\left[n_1^2 + (b-n_1)^2\right],
\end{equation}
one may obtain the energy dependence of the correlation function (\ref{eq:corrdimertheo}), which is plotted in Fig. \ref{corrdimer}-b (thin solid line). We can see that beyond the energy threshold for appearance of QBs, the numerical results are close to the estimation (\ref{eq:corrdimertheo}).

\subsection{Entanglement}

QBs may also be differentiated from other quantum states when measuring the degree of entanglement \cite{Tonel2005JPhysA38,Fu2006PRA74}. For the dimer and the trimer the degree of entanglement in the eigenstates may be measured by
minimizing the distance of a given state to the space of product states of the dimer part (expanded by the product basis $\{|n_1\rangle\otimes|n_2\rangle\}$), which depends on the largest eigenvalue of the corresponding reduced density matrix
\cite{Wei2003PRA68,Shimony1995AnnNYAcadSci755,Barnun2001JPhysA34,rapsf07arx}:
\begin{equation}
\Delta = \sum_{n_1,n_2}^N(\chi_{n_1,n_2} - f_{n_1}g_{n_2})^2,
\end{equation}
where for the case of the dimer $\chi_{n_1,n_2}=\langle n_1,n_2|\chi\rangle$,
and for the trimer $\chi_{n_1,n_2}=\langle n_1,n_2,(n_3=b-n_1-n_2)|\chi\rangle$.
The functions $f_{n_1}$ and $g_{n_2}$ are such that $\Delta$ is minimum \cite{rapsf07arx}.
$\Delta$ measures how far a given eigenstate of
the system is from being a product of single-site
states, and has values $0<\Delta<1$. This measure has a direct relation to the distance of a given eigenstate from a possible one obtained after performing a Hartree approximation \cite{Wei2003PRA68}.
\begin{figure}[!t]
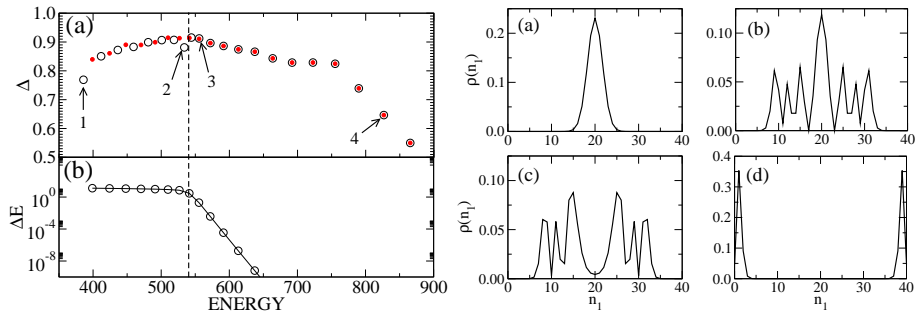

\begin{center}
\includegraphics[width=2.3in]{entanglement_splitting_b_40_c_2_2.eps}
\includegraphics[width=2.4in]{density_dimer_b_40_c_2_2.eps}
\caption{\label{entsplittdimer}Left panel: (a) Entanglement of the eigenstates and (b)
energy splitting as a function of energy in the dimer (open circles, symmetric
eigenstates; solid circles, antisymmetric
eigenstates). The vertical dashed line marks the energy threshold for appearance
of QB states. Here $b=40$, and $C=2$. Right panel: The density of the
symmetric eigenstates marked by labeled arrows in left panel-a:
(a) S-0 (arrow 1), (b) S-7 (arrow 2), (c) S-9 (arrow 3), (d) S-19 (arrow
4).}
\end{center}
\end{figure} 

For the dimer, since QB states are close
to eigenstates of the $C=0$ case 
\begin{equation}\label{eq:QB}
|\chi\rangle_{QB} \simeq \frac{1}{\sqrt{2}}(|n,0\rangle \pm |0,n\rangle ),
\end{equation}
with $n\lesssim b$, one expects that the degree of entanglement in QB
states is similar to the degree of entanglement in such states.
Since only two basis states are involved, it can not be a state of maximum
entanglement.
For $C=0$ the
eigenstates of the system are the basis states given by eq. (\ref{eq:dimerbasis}), where for
$n_1=n_2$ it follows that $\Delta=0$, and for $n_1\neq n_2$ (which includes the state
in eq. \ref{eq:QB}) $\Delta = 0.5$.

In previous works in a similar quantum dimer model \cite{Tonel2005JPhysA38,Fu2006PRA74}, it was
shown that at the energy
threshold for appearance of QB states the entanglement (in this case measured
in a different way) becomes maximum and then decreases with
energy. From this, and the above reasoning, we expect that QB states show decreasing
entanglement $\Delta$ with energy, tending to 0.5. Results in left panel-a of Fig. \ref{entsplittdimer} agree with this expectation.

For $C=0$, the entanglement has the values 0 and 0.5 corresponding to the basis states
$|b/2,b/2\rangle$ and $|n,b-n\rangle$ ($n\neq b/2$) with
equal and distinct number of quanta at each site respectively. 
When $C>0$, the eigenstates become linear superpositions of
the basis states and the entanglement rises, being larger as long as more
basis states are involved in building up an eigenstate. This can be seen in the right panel of
Fig. \ref{entsplittdimer}, where we plot the density $\rho(n_1,n_2=b-n_1)=|\langle
n_1,b-n_1|\chi\rangle|^2\equiv\rho(n_1)$ of four
symmetric eigenstates marked by labeled arrows in the left panel: The low-energy eigenstate marked by the arrow 1 consists mainly of one basis state: $|b/2,b/2\rangle$,
as seen in the right panel-a, hence the entanglement is
relatively small. When going up in energy the entanglement in the eigenstates
quickly increases, becoming maximum at the energy threshold,
and then decreases. An eigenstate just before the threshold like
the one marked by the arrow 2 in the left panel involves many
basis states fulfilling $n_1+n_2=b=40$ (right panel-b), hence the
entanglement is large. However, for a QB state lying in the energy region
beyond the threshold, like the one marked by the
arrow 3 in the left panel, the number of involved basis states,
and thus the entanglement starts to decrease (right panel-c). Finally,
in high-energy eigenstates like the one marked by the arrow 4 in the left panel, which has the form shown in eq. (\ref{eq:QB}) (right panel-d), the entanglement is even smaller and gets close to 0.5 as expected.

From the above results we see that by measuring entanglement one may gain information not only about the energy threshold for existence of QBs (also visible when measuring the energy splitting and correlation function), but also about how many basis states overlap strongly with the eigenstate under consideration. 
We also computed the {\it von Neumann entropy} \cite{Hines2003PRA67}, which is
another standard measure of entanglement, and the results were consistent with
those discussed above.

\section{Quantum edge-localized states}


Most of the studies about QBs in large lattices (with few bosons) were done considering a system with periodic boundary conditions, and thus, translational invariant. However, ususally real systems have to be modeled with open boundary conditions. Hence it is natural to wonder what hapens with QBs when the lattice has finite size, and therefore, no translational invariance.

In the classical case, it has been shown that in finite nonlinear lattices, the breaking of the translational symmetry may lead to the formation of so called {\it nonlinear edge states}. These are excitations which are localized at the edges of the lattice and they have been studied, in particular, in nonlinear optics experiments employing optical waveguides, being coined with the name of {\it discrete surface solitons} \cite{MakrisSuntsovChristodoulidesStegemanHache_OptLett2005}. 
In particular, it is well known that the discrete nonlinear Schr\"odinger equation (DNLS) has time-periodic solutions localized at the edge of the lattice \cite{MolinaVicencioKivshar_OptLett2006,Suntsov2006PRL96_and_others}. It is therefore expected that the large-boson limit of the open-boundary Bose-Hubbard model will show
eigenstates in which the bosons are localized at the edge.

\begin{figure}[!t]
\begin{center}
\includegraphics[width=3.in]{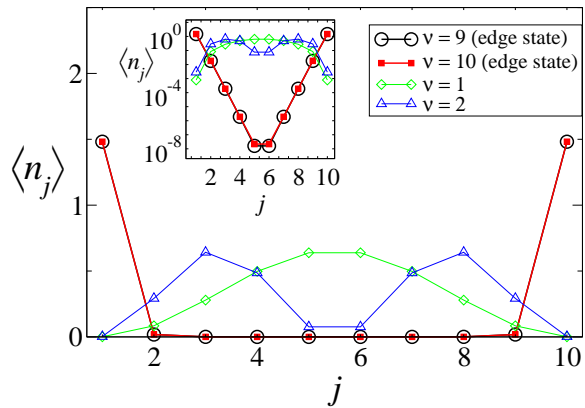}
\caption{\label{3bosonedgestates}
Spatial profile of site occupancies for several eigenstates
of the three-boson Bose-Hubbard chain. The index $\nu$ counts the eigenstates from the lowest-energy one ($\nu=1$). Inset shows the same plot in semilog
scale, the linear behavior indicating exponential localization in the edge
states. This image was taken from Ref. \cite{Pintoedge}}
\end{center}
\end{figure}

Numerical studies by Pouthier were done to answer the edge-localization question in a lattice with a few bosons \cite{Pouthier_PRB2007}, where the mean-field approximation (DNLS
equation) cannot \emph{a priori} be expected to provide the correct intuition. The answer turns out to be subtle --- this phenomenon is not present for the
case of two particles, but appears when the particle number is three or more \cite{Pouthier_PRB2007}. Further studies focused on the energy spectrum and eigenstates (Fig. \ref{3bosonedgestates}) were done by Pinto {\it et al} \cite{Pintoedge}, where degenerate perturbation theory allowed to explain why edge-localized states exist only if the number of particles is three or more \cite{Pintoedge}.

\section{Quantum breathers with fermions}

Advances in experimental techniques of manipulation of ultracold atoms in optical
lattices make it feasible to explore the physics of few-body interactions. Systems with few quantum particles
on lattices have new unexpected features as compared to the condensed matter case of
many-body interactions, where excitation energies are typically small compared to the Fermi energy.
Therefore it is of interest to study binding properties of fermionic pairs with total spin zero, as recently
presented in Ref.\cite{jpnsf09}.
We use the extended Hubbard model, which contains two interaction scales - the on site interaction $U$ and
the nearest neighbour intersite interaction $V$. The nonlocal interaction $V$ is added in condensed matter physics to emulate
remnants of the Coulomb interaction due to non-perfect screening of electronic charges.
For fermionic ultracold atoms or molecules with magnetic or electric 
dipole-dipole interactions, it can be tuned with respect to
the local interaction $U$ by modifying the trap geometry of a condensate, additional
external dc electric fields, combinations with fast rotating external fields, etc (for a review
and relevant references see \cite{mab08}).

Consider a one-dimensional lattice with
$f$ sites and periodic boundary conditions described by the extended Hubbard model with the following Hamiltonian:
\begin{equation}\label{eq:hamiltonian}
\hat{H} = \hat{H}_0 +  \hat{H}_U + \hat{H}_V,
\end{equation}
where
\begin{equation}
\hat{H}_0=-\sum_{j,\sigma} \hat{a}^+_{j,\sigma} (\hat{a}_{j-1,\sigma} + \hat{a}_{j+1,\sigma} )\;,
\end{equation}
\begin{equation}\label{eq:haminteraction1}
\hat{H}_U = -U\sum_{j} \hat{n}_{j,\uparrow} \hat{n}_{j,\downarrow}\;,\;\hat{n}_{j,\sigma}=
\hat{a}^+_{j,\sigma} \hat{a}_{j,\sigma}\;,
\end{equation}
\begin{equation}\label{eq:haminteraction2}
\hat{H}_V = -V\sum_{j} \hat{n}_j \hat{n}_{j+1}\;,\; \hat{n}_j = \hat{n}_{j,\uparrow}+ \hat{n}_{j,\downarrow}\;.  
\end{equation}
$\hat{H}_0$ describes the nearest-neighbor hopping of fermions
along the lattice. Here the symbols $\sigma = \uparrow,\downarrow$ stand for
a fermion with spin up or down.  $\hat{H}_U$
describes the onsite interaction between the particles, and 
$\hat{H}_V$  the intersite interaction of fermions located at adjacent sites.
$ \hat{a}^+_{j,\sigma}$ and $\hat{a}_{j,\sigma}$ are the fermionic creation and
annihilation operators satisfying the corresponding anticommutation relations: 
$\{\hat{a}^+_{j,\sigma},\hat{a}_{l,\sigma'}\}=\delta_{j,l}\delta_{\sigma,\sigma'}$,
$\{\hat{a}^+_{j,\sigma},\hat{a}^+_{l,\sigma'}\}=\{\hat{a}_{j,\sigma},\hat{a}_{l,\sigma'}\}=0$.
Note that throughout this work we consider $U$ and $V$ positive, which leads to bound states
located below the two-particle continuum. A change of the sign of $U,V$ will simply swap the energies.

To observe the fermionic character of the considered states, any two-particle number state
is generated from the vacuum $|O\rangle$ by first creating a particle with spin down,
and then a particle with spin up: e.g. $\hat{a}^+_{2,\uparrow} \hat{a}^+_{1,\downarrow} |O\rangle$
creates a particle with spin down on site 1 and one with spin up on site 2, while 
$\hat{a}^+_{2,\uparrow} \hat{a}^+_{2,\downarrow} |O\rangle$ creates both particles
with spin down and up on site 2.

Due to periodic boundary conditions the Hamiltonian (\ref{eq:hamiltonian}) commutes also with the 
translation operator $\hat{T}$, which shifts all lattice indices by one.
It has eigenvalues $\tau=exp(ik)$, with Bloch wave number $k=\frac{2\pi\nu}{f}$ and $\nu=0,1,2,...,f-1$.

For the case of having only one fermion (either spin up or spin down) in the lattice ($n=1$),
a number state has the form $|j\rangle = \hat{a}^+_{j,\sigma} |O\rangle$.
The interaction terms $\hat{H}_U$ and
$\hat{H}_V$  do not contribute.
For a given wave number $k$, the eigenstate to (\ref{eq:hamiltonian}) is therefore given by:
\begin{equation}
|\Psi_1\rangle=\frac{1}{\sqrt{f}} \sum_{s=1}^f\Big(\frac{\hat{T}}{\tau}\Big)^{s-1}|1\rangle\;.
\end{equation}
The corresponding eigenenergy 
\begin{equation}\label{eq:1fermionenergy}
\varepsilon_k=-2\cos(k).
\end{equation}

For two particles,  the number state method involves $N_{s}= f^2$  basis states,
which is the number of ways one can distribute two
fermions with opposite spins over the $f$ sites including possible double occupancy of a site.
Below we consider only cases of odd $f$ for simplicity. Extension to even values of $f$ is straightforward.
The details of the calculations can be found in \cite{jpnsf09}.

\begin{figure}
\begin{center}
\includegraphics[width=0.6\columnwidth]{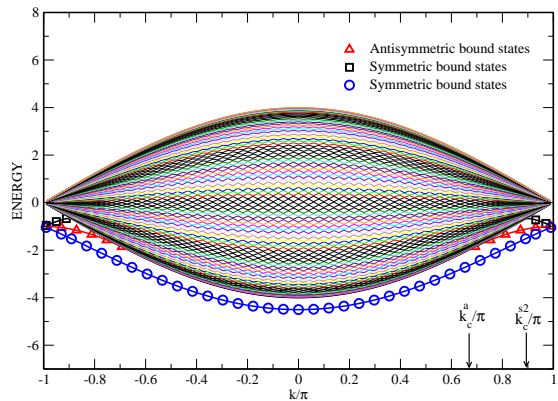}
\caption{\label{spectrum}
(Color online)
Energy spectrum of the two fermion states.
The eigenvalues are plotted as a function of the wave
number $k$. Here $U=1\;,\;V=1,\;f=101$. 
Symbols are
from analytical derivation, lines are the result of numerical diagonalization.
The arrows indicate the location of the critical wave numbers (see text).
Adapted from \cite{jpnsf09}.
}
\end{center}
\end{figure} 

In Fig. \ref{spectrum} we show the energy spectrum of the
Hamiltonian matrix obtained by numerical diagonalization for
the interaction parameters $U=2$ and  $V=2$ and $f=101$. 
At $U=0$ and $V=0$, the spectrum is given by the two fermion continuum, whose
eigenstates are
characterized by the two fermions independently moving along the
lattice. In this case the eigenenergies are the sum of the two single-particle energies:
\begin{equation}\label{eq:energyunperturbed}
E_{k_1,k_2}^0=-2[\cos(k_1)+\cos(k_2)],
\end{equation}
with $k_{1,2}={\pi}\nu_{1,2}/(f+1)$ , 
$\nu_{1,2}= 1,\ldots,f$. 
The Bloch wave number $k = k_1+k_2 \mod 2\pi$. Therefore, if $k=\pm \pi$,
the continuum degenerates into points.
The continuum is bounded by the hull curves $h_{\pm}(k)=\pm 4 \cos \frac{k}{2}$.
The same two-particle continuum is still observed in 
Fig. \ref{spectrum} for nonzero interaction.
However, in addition to the continuum, we observe one, two or three bound states
dropping out of the continuum, which depends on the wave number.
For any nonzero $U$ and $V$, all three bound states drop out of the 
continuum at $k=\pm \pi$.
One of them stays bounded for all values of $k$. The two other ones merge with the
continuum at some critical value of $|k|$ upon approaching $k=0$ as observed in
Fig.\ref{spectrum}. Note that for $k=\pm \pi$ and $U=V$, all three bound states
are degenerate.

\begin{figure}
\begin{center}
\includegraphics[width=0.6\columnwidth]{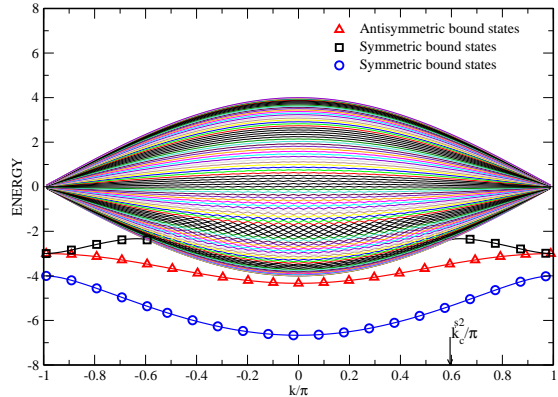}
\caption{\label{pbcspectrum}
(Color online)
Energy spectrum of the two fermion states.  
The eigenvalues are plotted as a function of the wave
number k.  Here $U=4\;,\;V=3\;,\;f=101$. The symbols are
from analytical derivation, lines are the results of numerical diagonalization.
The arrow indicates the location of the critical wave number (see text).
Adapted from \cite{jpnsf09}.}
\end{center}
\end{figure}

Upon increasing $U$ and $V$, we observe
that a second bound state band separates from the continuum for all $k$ (Fig.\ref{pbcspectrum}). At the same time,
when $U \neq V$,
the degeneracy at $k=\pm \pi$ is reduced to two.
 
\begin{figure}
\begin{center}
\includegraphics[width=0.6\columnwidth]{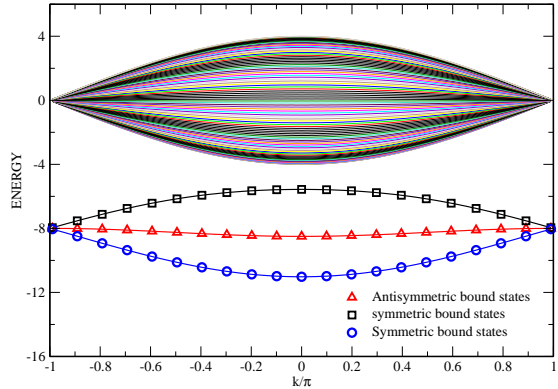}
\caption{\label{symspect} 
(Color online)
Energy spectrum for the two fermion states.
The eigenvalues are plotted as a function of the wave
number k.  Here $U=8\;,\;V=8\;,\;f=101$. The symbols are
from analytical derivation, lines are the results of numerical diagonalization.
Adapted from \cite{jpnsf09}.
}
\end{center}
\end{figure}

Finally, for even larger values of $U$ and $V$, all three bound state bands completely
separate from the continuum (Fig.\ref{symspect}).

\subsection{ Symmetric and antisymmetric state representation}
In order to obtain analytical estimates on the properties of the observed bound states,
we use the fact that the Hamiltonian for a two fermion state is invariant under flipping
the spins of both particles. 
We define symmetric basis states
\begin{equation}\label{eq:basisym}
|\Phi_{j,s}\rangle=\frac{1}{\sqrt{2}}(|\Phi_{j,+}\rangle + |\Phi_{j,-}\rangle)
\end{equation}
and antisymmetric states
\begin{equation}\label{eq:basisas}
|\Phi_{j,a}\rangle=\frac{1}{\sqrt{2}}(|\Phi_{j,+}\rangle - |\Phi_{j,-}\rangle)\;.
\end{equation}\label{fig8-6-1}

Note that $|\Phi_1\rangle$ is a symmetric state as well.


\subsection{ Antisymmetric bound states}
The antisymmetric states exclude double occupation. Therefore the spectrum is identical
with the one of two spinless fermions \cite{acsjcehg94}.
Following the derivations in \cite{acsjcehg94}
we find that the antisymmetric bound state, if it exists, has an energy
\begin{equation}
\label{eanti}
E^a_2(k)=-(V+\frac{4}{V}\cos^2(\frac{k}{2}))\;.
\end{equation}
This result is valid as long as the bound state energy stays outside of the 
continuum. The critical value of $k$ at which validity is lost, is obtained
by requesting $|E^a_2(k)|=|h_{\pm}(k)|$. 
It follows
$ V=2\cos(\frac{k}{2})$. 
Therefore the antisymmetric bound state merges with the continuum at a critical wave number
\begin{equation}
\label{kac}
 k^a_{c}=2\arccos(\frac{V}{2})\;,
\end{equation} 
setting a critical length scale $\lambda^a_{c}=\frac{2\pi}{ k^a_{c}}$.
For $V=1$ it follows $k^a_{c}/\pi\approx 0.667$ (see Fig.\ref{spectrum}).

The equation (\ref{eanti}) is in excellent agreement with the numerical data in 
Figs. \ref{spectrum},\ref{pbcspectrum},\ref{symspect} (cf. open triangles). We also note, that the
antisymmetric bound state is located between the two symmetric bound states, which we
discuss next. 

\subsection{Symmetric bound states}
A bound state can be searched for by assuming an unnormalized eigenvector of the form
$|c,1,\mu,\mu^2,\mu^3,...\rangle$ with $|\mu| \equiv \rho \leq 1$.
We obtain \cite{jpnsf09}
\begin{equation}{\label{eq:ensym}}
E^s_2(k) = -2({\rho+\frac{1}{\rho})\cos{k/2}}\;.
\end{equation}
The parameter $\rho$ satisfies a cubic equation
\begin{equation}\label{eq:cubiceq}
 a{\rho}^3+b{\rho}^2+c{\rho}+d=0
\end{equation}
with the real coefficients a, b. c and d given by
$a=2Vcos(\frac{k}{2})$, 
$b= 4cos^2(\frac{k}{2})-UV$,
$c=2(U+V)cos(\frac{k}{2})$,
$d=-4cos^2({\frac{k}{2}})$.
We plot the results in Figs. \ref{spectrum},\ref{pbcspectrum},\ref{symspect} (cf. open circles
and squares). We obtain excellent agreement. 

At the Brilloin zone edge $k=\pm \pi$ the cubic equation (\ref{eq:cubiceq})
is reduced to a quadratic one, and can be solved to obtain finally $\rho \rightarrow 0$ and
\begin{equation}{\label{eq:enrhosmal}}
E^{s1}_{2}(k\to\pm\pi)=-U\;,\;E^{s2}_{2}(k\to\pm\pi) =-V \;.
\end{equation}
In particular we find for $k=\pm \pi$ that $E^{s2}_{2} = E^a_2$. In addition, if $U=V$,
all three bound states degenerate at the zone edge.

If $V=0$, the cubic equation (\ref{eq:cubiceq}) is reduced to a quadratic one in the whole
range of $k$ and yields \cite{acsjcehg94}
\begin{equation}
E^{s1}_{2}(k) = -\sqrt{U^2 + 16 \cos^2 ( k/2)}\;.
\end{equation}

Next we determine the critical value of $k$ for which the bound state with energy $E^{s2}_{2}$
is joining the continuum. Since at this point $\rho=1$, we solve (\ref{eq:cubiceq})
with respect to $k_c$ and find
\begin{equation}
\label{ks2c}
k^{s2}_{c}=2\arccos\Big( \frac{UV}{2(U+2V)}\Big)
\end{equation}
setting another critical length scale $\lambda^s_{c}=\frac{2\pi}{ k^s_{c}}$.
E.g. for $U=V=1$ $k^{s2}_{c}/\pi \approx 0.89$, in excellent agreement with Fig.\ref{spectrum}.
For $U=4$ and $V=3$ we find $k^{s2}_{c}/\pi \approx 0.59$ confirming numerical results in Fig.\ref{pbcspectrum}.

\section{Small Josephson junction networks}

Recent studies of Pinto et al \cite{rapsf07,rapsf07arx} deal with quantum breather
excitations in two capacitively coupled Josephson junctions. Such systems
are currently under experimental investigation, being candidates for quantum
information processing, and show remarkably long coherence times up to
100 ns for few quanta excitations. The system does not conserve the number of excited quanta,
and can be best compared with the above Bose-Hubbard trimer. Quantum breather
signatures are found simultaneously in the spectra (tunneling splittings),
correlation functions, entanglement, and quanta number fluctuations.

We address the excitation of QBs in a system of two coupled Josephson junctions in the
quantum regime \cite{rapsf07arx}. Josephson junctions are nonlinear devices that show macroscopic quantum behavior, and nowadays they can be manipulated with high precision, in such a way that the energy flow between coupled junctions can be resolved in time.

Josephson junctions behave like anharmonic
oscillators, and by lowering the temperature one can bring them into the
quantum regime, where their quantization leads to energy
levels which are nonequidistant because of the anharmonicity. These levels can be
separately excited by using microwaves pulses, and the energy distribution
between the junctions can be measured in time using subsequent
pulses \cite{MacDermott2005Science307}. So far these techniques have been used for experiments on quantum information processing with Josephson junctions \cite{Leggett,Esteve,You2005PhysTodayNov}, but we think that arrays of Josephson junctions in the qunatum regime also might be used as playgrounds for experiments on quantum dynamics of excitations in nonlinear lattices.

\begin{figure}[!t]
\begin{center}
\includegraphics[width=0.2\columnwidth]{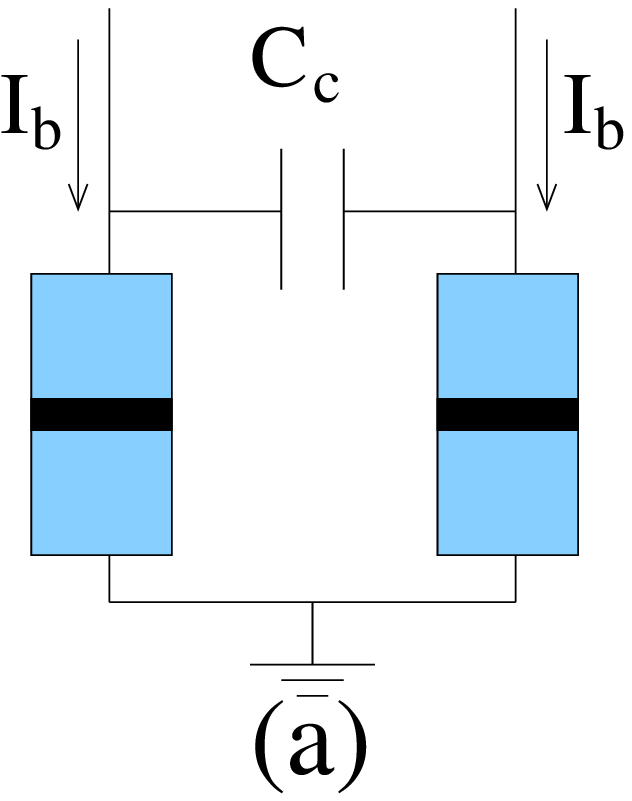}
\hspace*{5mm}
\includegraphics[width=0.2\columnwidth]{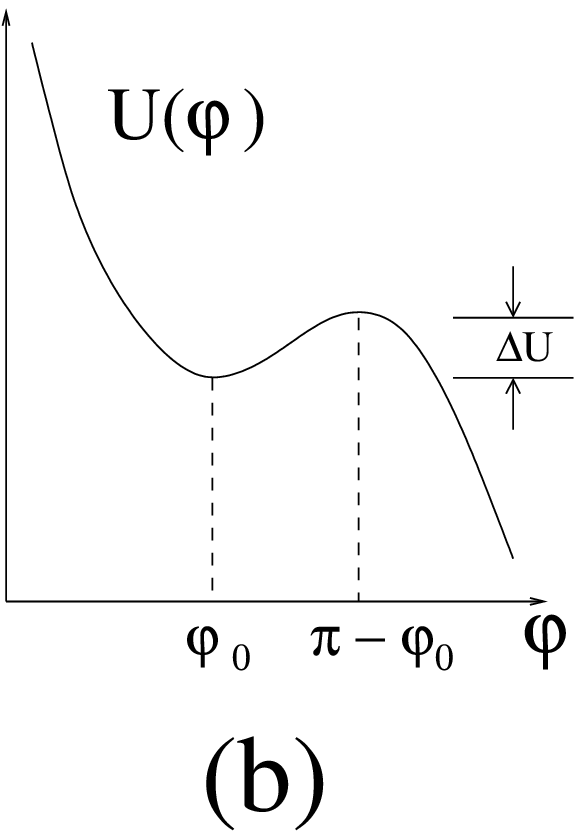}
\hspace*{5mm}
\includegraphics[width=0.3\columnwidth]{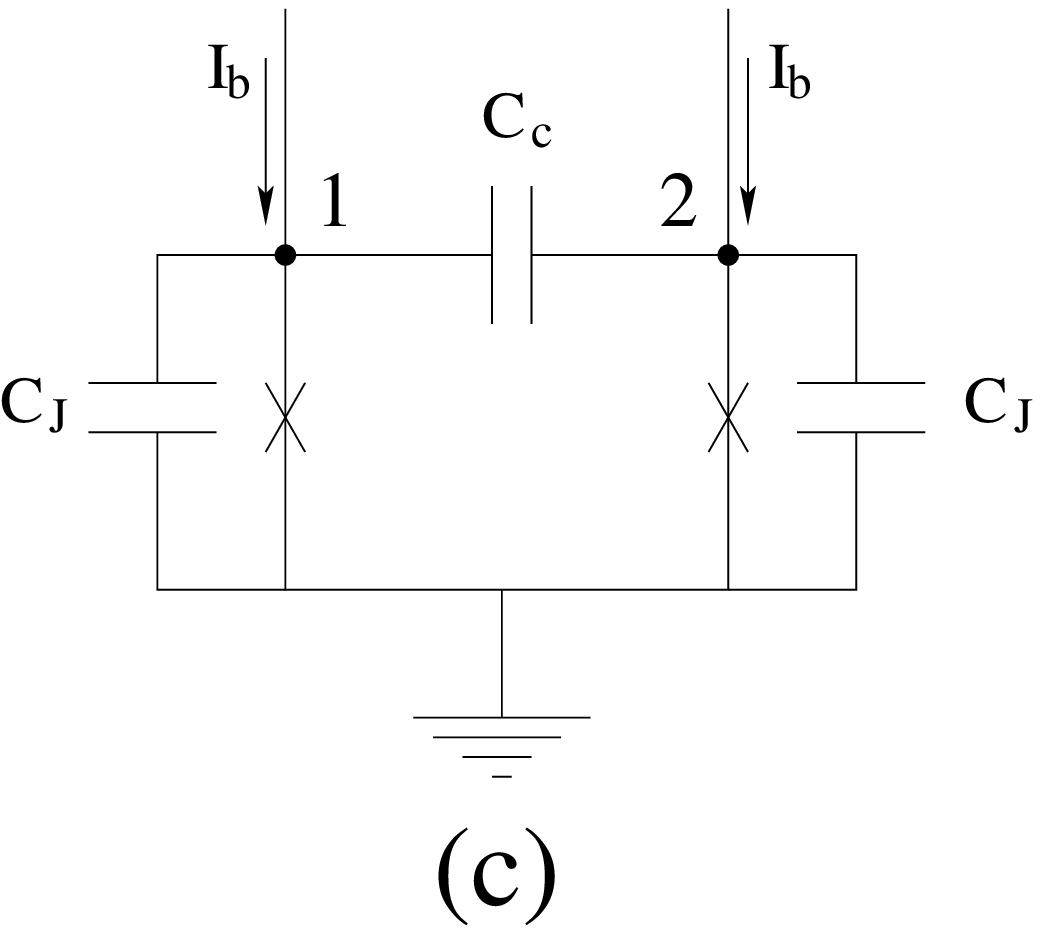}
\caption{\label{fig1}(a) Sketch of the two capacitively coupled Josephson
junctions. (b) Sketch of the washboard potential for a single
current-biased Josephson-junction. (c) Circuit diagram for two ideal capacitively coupled
Josephson junctions.}
\end{center}
\end{figure}

The system is sketched in Fig.\ref{fig1}-a: two JJs are coupled by a
capacitance $C_c$, and they are biased by the same current $I_b$. The strength
of the coupling due to the capacitor is $\zeta =
C_c/(C_c+C_J)$.
The dynamics of a biased
Josephson junction (JJ) is analogous to the dynamics of a particle with a mass
proportional to the junction capacitance $C_J$,
moving on a tilted washboard potential
\begin{equation}
U(\varphi) = -I_c\frac{\Phi_0}{2\pi}\cos\varphi - I_b\varphi\frac{\Phi_0}{2\pi},
\end{equation}
which is sketched in Fig.\ref{fig1}-b.
Here $\varphi$ is the phase difference between the macroscopic wave functions
in both superconducting electrodes of the
junction, $I_b$ is the bias current, $I_c$ is the critical current of the junction, and $\Phi_0=h/2e$ the
flux quantum. When the energy of the particle is large enough to overcome the barrier
$\Delta U$ (that depends on the bias current $I_b$) it escapes and moves down the
potential, switching the junction into a resistive state with a nonzero voltage
proportional to $\dot{\varphi}$ across it.
Quantization of the system leads to discrete energy levels inside the
potential wells, which are nonequidistant because of the anharmonicity. Note that even if there is not
enough energy to classically overcome the barrier, the
particle may perform a quantum escape and tunnel outside the well, thus switching the junction into the resistive
state. Thus each state inside the well is characterized by a bias and state-dependent
lifetime, or its inverse ---the escape rate.

The Hamiltonian of the system is
\begin{equation}
H = \frac{P_1^2}{2m}+ \frac{P_2^2}{2m} + U(\varphi_1)+U(\varphi_2) + \frac{\zeta}{m}P_1P_2,
\end{equation}
where
\begin{eqnarray}
m &=& C_J(1+\zeta)\left(\frac{\Phi_0}{2\pi}\right)^2 ,\\
P_{1,2} &=& (C_c+C_J)\left(\frac{\Phi_0}{2\pi}\right)^2 (\dot{\varphi}_{1,2} -
\zeta\dot{\varphi}_{2,1}).
\end{eqnarray}
Note that the conjugate momenta $P_{1,2}$ are proportional to the charge at
the nodes of the circuit (which are labeled in Fig.\ref{fig1}-c).

\begin{figure}[!t]
\begin{center}
\includegraphics[width=0.6\columnwidth]{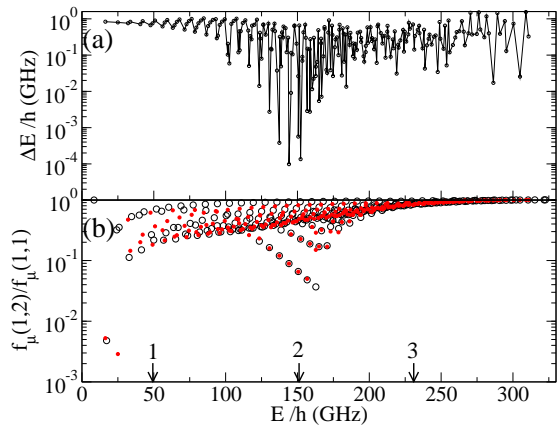}
\caption{\label{fig2}(a) Energy splitting and (b) correlation function
vs. energy of the eigenstates of the two-junctions system (open circles, symmetric
eigenstates; filled circles, antisymmetric eigenstates). The labeled arrows mark the energy
corresponding to the peak of the spectral intensity in Fig.\ref{fig3}-b, d, and f (see text).
The parameters are
$\gamma=Ib/Ic=0.945$ and $\zeta=0.1$ (22 levels per junction).}
\end{center}
\end{figure}

In the quantum case the energy eigenvalues and the eigenstates of the system 
were computed and analyzed in \cite{rapsf07arx}.
In Fig.\ref{fig2} we show the nearest neighbor energy spacing 
(tunneling splitting) and the correlation function of
the eigenstates. For this, and all the rest, we used
$I_c=13.3 \;\mu$A, $C_J=4.3$ pF, and $\zeta=0.1$, which are typical values in experiments.
We see that in the central part of the spectrum the energy splitting becomes
small in comparison to the average. The corresponding pairs of eigenstates, which are tunneling pairs,
are site correlated, and thus QBs. In these states
many quanta are localized on one junction and the tunneling time of
such an excitation from one junction to the other (given by the inverse energy
splitting between the eigenstates of the pair) can be exponentially large and depend
sensitively on the number of quanta excited.

Note that the tunneling of quanta between the JJs occurs without an obvious potential energy barrier being present (the interaction between the junctions is only through their momenta). This process has been coined {\it dynamical tunneling} \cite{Davis1981JChemPhys75,Keshavamurthy2007IntRevPhysChem26,Keshavamurthy2005JChemPhys122}, to distinguish from the usual tunneling through a potential barrier. In dynamical tunneling, the barrier ---a so-called invariant separatrix manifold--- is only visible in phase space, where it separates two regions of regular classical motion between which the tunneling process takes place. Therefore, when referring to the tunneling between the JJs, we implicitly mean that it is dynamical.

The fact that the strongest site correlated eigenstates occur in the
central part of the energy spectrum may be easily explained as follows: Let $N$ be
the highest excited state in a single junction, with a corresponding
maximum energy $\Delta U$ (Fig.\ref{fig1}). For
two junctions the energy of the system with both junctions in the
$N$-th state is $2\Delta U$, which roughly is the width of the full
spectrum. Thus states of the form $|N,0\rangle$ and $|0,N\rangle$ that
have energy $\Delta U$ are located approximately in the middle.

\begin{figure}[!t]
\begin{center}
\includegraphics[width=0.5\columnwidth]{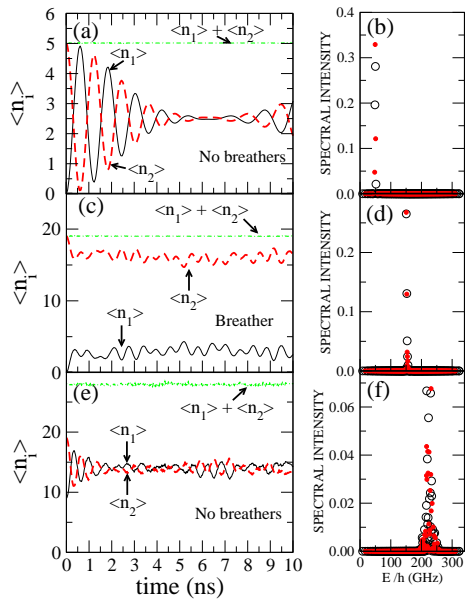}
\caption{\label{fig3}Time evolution of expectation values of the number of
quanta at each junction (left panels) for different initial excitations with corresponding spectral intensities (right panels). (a) and (b): $|\Psi_0\rangle =|0,5\rangle$;
(c) and (d): $|\Psi_0\rangle =|0,19\rangle$;
(e) and (f): $|\Psi_0\rangle =|9,19\rangle$. Open circles, symmetric
eigenstates; filled circles, antisymmetric eigenstates. The energies of the peaks in the spectral
intensity are marked
by labeled arrows in Fig.\ref{fig2}-b (see text). The parameters are
$\gamma=Ib/Ic=0.945$ and $\zeta=0.1$ (22 levels per junction).}
\end{center}
\end{figure} 

Having the eigenvalues and eigenstates, we compute the time evolution of
different initially localized excitations, and the expectation value of
the number of quanta at each junction $\langle \hat{n}_i\rangle (t)=
\langle\Psi(t)|\hat{n}_i|\Psi(t)\rangle$. Results are shown in
Fig.\ref{fig3}a, c, and e. We also compute the spectral intensity
$I_{\mu}^0 = |\langle \chi_{\mu}|\Psi_0\rangle|^2$, which
measures the strength of overlap of the initial state $|\Psi_0\rangle$ with the
eigenstates.
Results are shown in Fig.\ref{fig3}-b,
d, and f, where we can see a peak in each case, which corresponds to the 
arrows in Fig.\ref{fig2}-b.
We
can see that the initial state $|\Psi_0\rangle=|0,5\rangle$ overlaps with
eigenstates with an energy
splitting between them being relatively large and hence the
tunneling time of the initially localized excitation is short.
For the case $|\Psi_0\rangle=|0,19\rangle$ QBs are excited: The
excitation overlaps strongly with tunneling pairs of eigenstates in the
central part of the spectrum, which are
site correlated and nearly degenerate. The
tunneling time of such an excitation is very long, and thus keeps the quanta localized 
on their initial excitation site for corresponding
times.
Finally the initial state $|\Psi_0\rangle = |9,19\rangle$ overlaps with weakly site correlated
eigenstates with large energy
splitting. Hence the tunneling time is short again.
\begin{figure}[!t]
\begin{center}
\includegraphics[width=0.5\columnwidth]{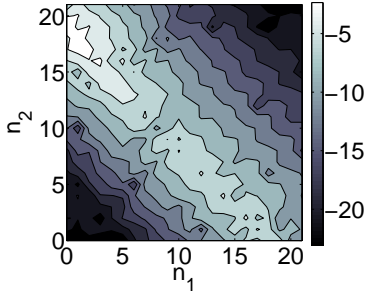}
\caption{\label{dist1}Contour plot of the logarithm of the density of the asymmetric state $|\chi\rangle =
(|\chi_{b}^{(S)}\rangle + |\chi_{b}^{(A)}\rangle)/\sqrt{2}$ as a function
of the number of quanta at junctions 1 and 2 (see text). The parameters are
$\gamma=Ib/Ic=0.945$ and $\zeta=0.1$ (22 levels per junction).}
\end{center}
\end{figure} 

We computed also the time evolution of the expectation values of the number of
quanta for initial conditions which are coherent or incoherent (mixtures) superpositions
of product basis states with equal weights. This is relevant for experiments,
since it may be hard to excite one junction
to a determined state but easier to excite several states of the junction at the
same time. 
We used coherent superpositions (characterized by well defined 	states
$|\Psi_0\rangle$), and mixtures (characterized by their corresponding density
operators $\hat{\rho}_0$), of four basis states around the
already used initial states: For the 
state $|0,5\rangle$ we superposed the basis
states $|0,5\rangle$, $|0,6\rangle$, $|0,7\rangle$, and $|0,8\rangle$,
for $|0,19\rangle$ the
basis states $|0,20\rangle$, $|0,19\rangle$, $|0,18\rangle$, and $|0,17\rangle$, and
for $|9,19\rangle$ the
basis states $|9,20\rangle$, $|9,19\rangle$, $|9,18\rangle$, and $|9,17\rangle$.
Both for superposition and mixture of basis states, the results
are qualitatively similar to those shown in Fig.\ref{fig3}. Therefore we
expect that some imprecision in exciting an initial state in the junctions
would not affect in a relevant way the results. we may also conclude, that
the excitation of QB states does not rely on the phase coherence.

Let us estimate how many quanta should be excited in the junctions in order to obtain
QBs (tunneling pairs).  We compute the density
$\rho(n_1,n_2)=|\langle n_1,n_2|\chi\rangle|^2$ of the asymmetric state  $|\chi\rangle =
(|\chi_{b}^{(S)}\rangle + |\chi_{b}^{(A)}\rangle)/\sqrt{2}$, where 
$|\chi_{b}^{(S,A)}\rangle$ are the eigenstates belonging to a tunneling
pair \cite{rapsf07arx}.
In Fig. \ref{dist1} we show a contour plot of the logarithm of the density for the 
tunneling pair with energy
marked by the arrow labeled by number two in Fig. \ref{fig2}-(b). We see
that the density has its maximum around $n_1=19$ and $n_2=0$ which is consistent
with the result shown in Fig. \ref{fig3}-c and d where QBs were excited
by using this combination of quanta in the junctions.

\section{Experimental realizations}

There is a fast growing amount of experimental and
related theoretical work on applying the quantum discrete breather
concept to many different branches in physics, like
Bose-Einstein condensates, crystals and molecules, surfaces,
and others. We will discuss some
of these at length, while others will be reviewed more briefly.

\subsection{Repulsively bound atom pairs}

\begin{figure}[!t]
\begin{center}
\includegraphics[angle=-0,width=0.5\textwidth]{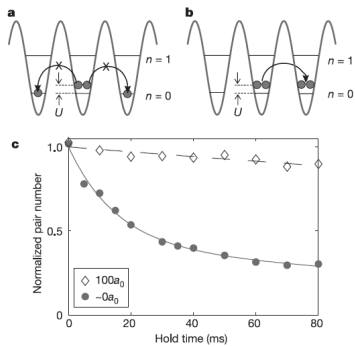}
\caption{\label{fig8-3-7}(a) Repulsive interaction between two atoms sharing a lattice site gives
rise to an interaction energy $U$. Breaking up of the pair is suppressed
owing to the lattice band structure and energy conservation.
This is the simplest version of a quantum discrete breather.
(b) The discrete breather can tunnel along the lattice.
(c) Long lifetimes of strongly repulsive atoms. The plot shows the remaining
fraction of pairs for strong interaction (open diamonds) and
for almost vanishing interaction (filled circles). 
Adapted from \cite{kwgtfl06}.}
\end{center}
\end{figure}

Winkler et al \cite{kwgtfl06} performed experiments with a three-dimensional
optical lattice with initially each site being either not occupied,
or being occupied by two Rb atoms bound in a pair due to attractive interaction.
A magnetic field sweep across the Feshbach resonance changes the sign of interaction,
turning attraction into repulsion. The dynamics of ultracold atoms loaded into
the lowest band of the optical potential is described by the quantum DNLS model,
which is equivalent to the Bose-Hubbard model (\ref{7-1}).
Lifetime measurements have shown, that repulsive pairs of Rb atoms have 
larger lifetimes than pairs of weakly or almost not interacting atoms
(Fig.\ref{fig8-3-7}).
\begin{figure}[!t]
\begin{center}
\includegraphics[angle=-0,width=0.6\textwidth]{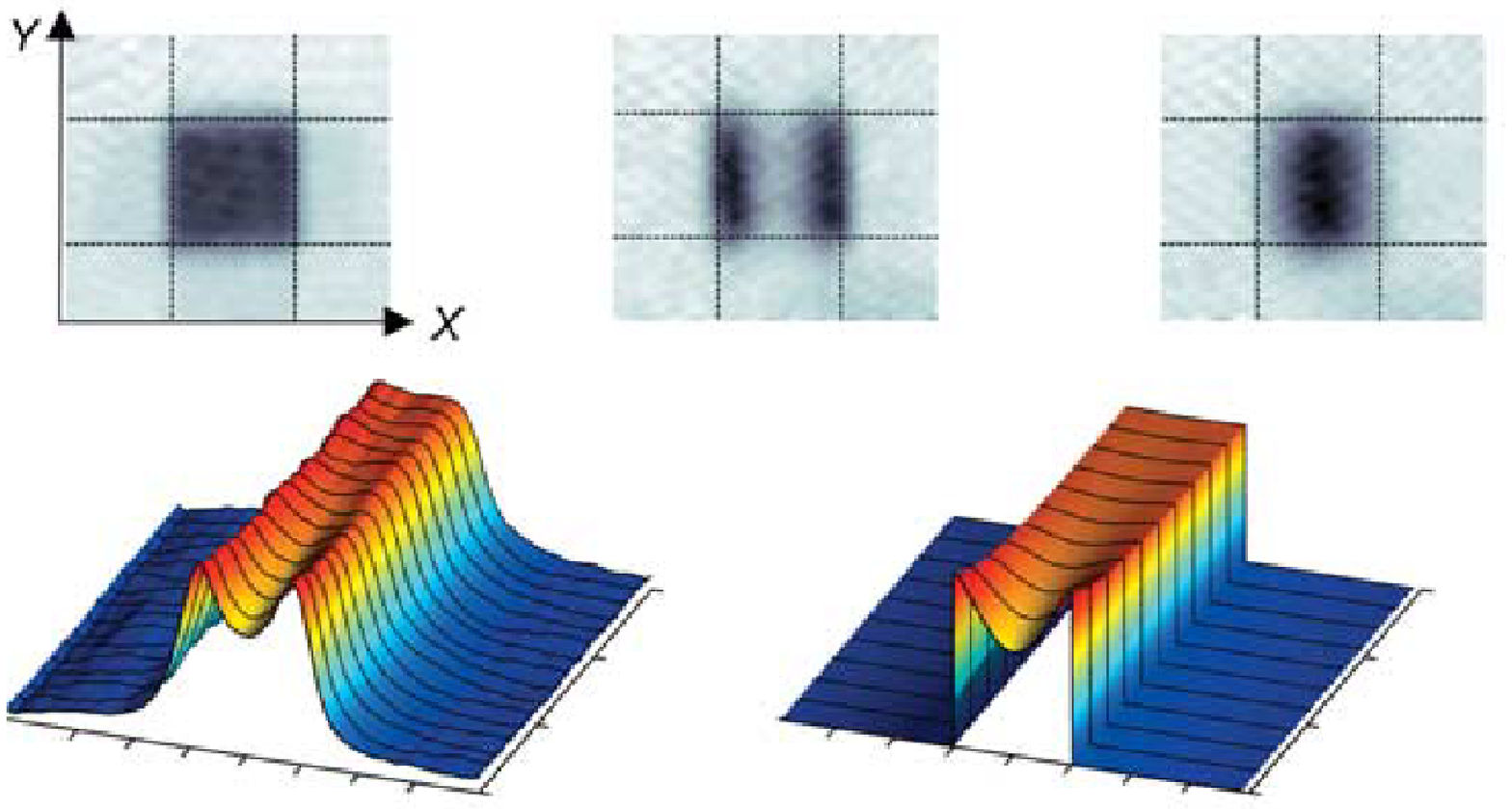}
\caption{\label{fig8-3-9}(Upper row) Absorption images of the atomic distribution after release from the
3D lattice and a subsequent 15 ms time of flight. The horizontal
and vertical black lines enclose the first Brillouin zone.
(upper left) Lattice sites are occupied by single atoms; (upper middle) Repulsively
bound atom pairs; (upper right) Attractively bound atom pairs;
(bottom row) Quasi-momentum distribution for pairs in one direction 
as a function of the lattice depth after integrating out the other
direction. (bottom left) Experiment; (bottom right) Numerical calculations.
Adapted from \cite{kwgtfl06}.}
\end{center}
\end{figure}
The two-particle bound states discussed in chapter \ref{sec7.1} -
the simplest versions of a quantum discrete breather - are the obvious
explanation of the experimental findings. Indeed, neglecting Landau-Zener transitions
to higher lying bands in the optical potential, the Bose-Hubbard model is justified.
The sign of the interaction does not play any role, since it only changes quantum discrete
breathers from being low-lying to being excited states, not affecting their localization properties.
The most simple argument of why two quanta (or atoms) placed initially close to each other,
do not separate despite they repel each other, is based on the fact, that 
if they would do so, the (large) interaction energy should be converted into kinetic
energy, which is restricted to be less than two times the width of the single
particle band. In other words, repulsively bound atom pairs are a straightforward
consequence of quantum discrete breather states with two quanta.

Another sophisticated experimental investigation aimed at measuring
the quasi-momentum distribution of atom pairs in various regimes by mapping it onto
a spatial distribution, which was finally measured using standard
absorption imaging (Fig.\ref{fig8-3-9}).
Therefore predictions of such states, which were made more than 30 years ago
by Ovchinnikov \cite{aao70}, were beautifully confirmed experimentally
with ultracold repulsive atoms.

\subsection{Molecules}

Intramolecular vibrational energy redistribution
(IVR) has been a central issue in the field of chemical physics
for many decades. In particular, pathways and rates are of importance there,
since understanding them allows 
to describe e.g. the dynamics of various chemical reactions, and dissociation
processes \cite{mjsygjbrsscf05}. Spectroscopical studies, where single 
vibrational quanta are excited, allow to measure the frequencies
of normal vibrational modes, i.e. to characterize the dynamics
of a molecule for small amplitude vibrations. 
These normal modes consist of coherent combinations of vibrational
excitations of several bonds (or rotational groups) in a molecule.
However, in order e.g.
to dissociate a molecule, a many quanta excitation is needed, and nonlinearities
will certainly become important. It was realized then, that
strong vibrational excitations of molecules are much better
described by so-called {\sl local modes}, i.e. basically
one or few bond vibrational excitations. That transition from
normal to local modes remained a puzzle for a long time.
A practically complete modern theoretical account on these issues
can be found in a recent monography by Ovchinnikov, Erikhman and Pronin
\cite{aaonsekap01}. 
On its most abstract level, the transition
from normal to local modes is identical with the bifurcation
in the dimer model.
Thus, local modes are essentially discrete breathers or slight perturbations
of them. Note, that the connection between local modes,
 breathers and periodic orbits
has been recently studied by Farantos in the context of large biological
molecules \cite{scf07}. Discrete breathers (ILMs) have been theoretically predicted
to exist in ionic crystals \cite{KSC1998}, ways of optical excitation of DBs (ILMs)
have been proposed \cite{RP1996}, and their possible presence in hydrocarbon structures
has been discussed \cite{KA2001}.

Exciting local modes in molecules with discrete symmetries
leads to small tunneling splittings of excitation levels \cite{aaonsekap01},
and goes back to the work of Child and Lawton \cite{mscrtl82},
see also a recent comprehensive review by Keshavamurthy \cite{sk07}
and references therein. On its most abstract level, this effect is
identical with the tunneling splitting in the permutationally
symmetric dimer model discussed in chapter \ref{sec7.2}.
\begin{figure}[!t]
\begin{center}
\includegraphics[angle=0,width=0.7\textwidth]{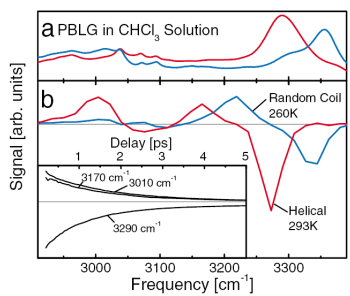}
\caption{\label{fig8-6-1}(a) Absorption spectra of PBLG in chloroform at 293 K (red line,
helical conformation) and at 260 K (blue line, random coil).
(b) Pump-probe spectra 600 fs after excita\label{fig8-6-2}
tion under the same conditions.
Inset: Decay of negative and both positive bands at 293 K.
Adapted from \cite{jerpvpcfph04}.}
\end{center}
\end{figure}

An early example of experimental evidence of discrete breather excitations
in molecules comes from spectroscopical studies of visible
red absorption spectra of benzene, naphtalene, and anthracene
by Swofford et al \cite{rlsmelaca76}. The C-H stretching vibrations
have been excited to the sixth quantum level, and red shifts of the
lines show, that instead of a delocalized excitation of six bonds to
the first level (yielding six quanta), the excitation resides on just
one of the six available bonds. While it can tunnel (as a quantum discrete
breather) to the other bonds, this tunneling time is a new large
time scale in the problem, strongly affecting e.g. dissociation rates.

A recent study of femtosecond infrared pump-probe spectroscopy
of the N-H mode of a stable $\alpha$-helix (poly-$\gamma$-benzyl-L-glutamate
(PBLG))
revealed two excited-state absorption bands, which disappear upon unfolding 
of the helix \cite{jerpvpcfph04}. PBLG forms extremely stable,
long $\alpha$-helices in both helicogenic solvents and films grown
from these solvents. The monomeric unit of PBLG is a non-natural
amino-acid with a long side chain that stabilizes the helix. PBLG has served 
as the standard model helix since the very early days of structural
investigations of proteins.
Fig.\ref{fig8-6-1}(a) (red line) shows the absorption spectrum
of the helix at 293 K, which is dominated by the strong
N-H stretching band at 3290 cm$^{-1}$. Fig.\ref{fig8-6-1}(b) (red line)
shows the pump-probe response 600 fs after excitation with an ultrashort
broadband pulse. One negative (3280 cm$^{-1}$) and two positive bands
(3160 and 3005 cm$^{-1}$) are observed. If the N-H stretching modes
were isolated, a negative band associated with bleach and stimulated emission, 
and a positive band associated with excited-state absorption, would be expected.
This is indeed observed here for the unfolded molecule.
In contrast, the observation of {\sl two} positive bands for the intact helix
rather than just one, is exceptional. Edler et al \cite{jerpvpcfph04}
\begin{figure}[!t]
\begin{center}
\includegraphics[angle=0,width=0.6\textwidth]{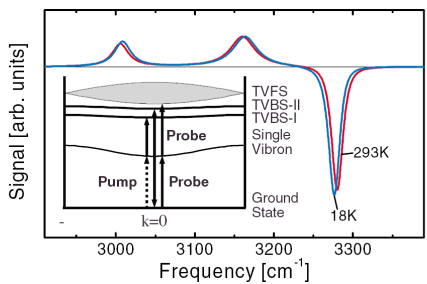}
\caption{\label{fig8-6-2}Simulated pump-probe spectrum for 293 K (red line) and 18 K (blue line).
Inset: schematic of the energy levels.
Adapted from \cite{jerpvpcfph04}.}
\end{center}
\end{figure}
argue, that these features can not be explained due to intensity dependencies, or
Fermi resonances. A consistent explanation is reached by assuming that
two vibron states are excited, and these vibrons may form two different
types of bound states, self trapped either on the same site, or on neighbouring
ones. The latter states originate from the acoustic phonons of the helix,
which correlate adjacent sites (see also \cite{phjeelt04}).

\subsection{Crystal surfaces}

Depositing atoms or molecules on crystal surfaces can be controlled experimentally,
and as a result a planar regular two-dimensional lattice structure
of the deposited material can be obtained. Guyot-Sionnest \cite{pgs91}
used Hydrogen to be deposited on Si(111) surfaces. The Si-H bonds
can be excited using pum-probe techniques with infrared dye lasers.
There is substantial interaction between the Si-H bonds on the Si(111)
surface. The pump excites one phonon (quantum), while the tunable probe frequency
finds a substantial red shift of the two-phonon excitation, and allows
to conclude, that two-phonon bound states are observed.

Another set of experiments was performed by Jakob \cite{pj96,pj98,pj99}.
Carbon monoxide (CO) was deposited on a Ru(001) single crystal surface.
The C-O stretching modes constitute a two-dimensional array of weakly interacting
anharmonic oscillators with 4.7 \AA ~intermolecular distance. 
Intermolecular coupling is provided by means
of the electric field of the oscillating dipoles.
\begin{figure}[!t]
\begin{center}
\includegraphics[angle=0,width=0.6\textwidth]{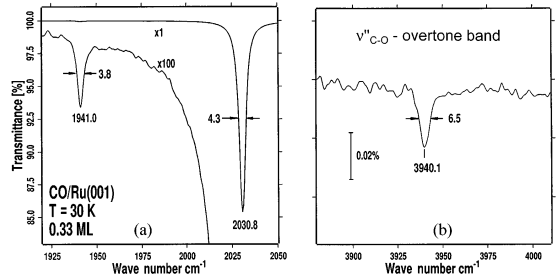}
\caption{\label{fig8-6-3}(a) Infrared absorption spectra of the C-O stretching mode at 30 K.
The corresponding mode of naturally abundant $^{13}$C$^{16}$O is displayed in 
an enlarged vertical scale; (b) the overtone band observed at less than twice
the frequency of the fundamental mode.
Adapted from \cite{pj96}.}
\end{center}
\end{figure}
\begin{figure}[!t]
\begin{center}
\includegraphics[angle=0,width=0.5\textwidth]{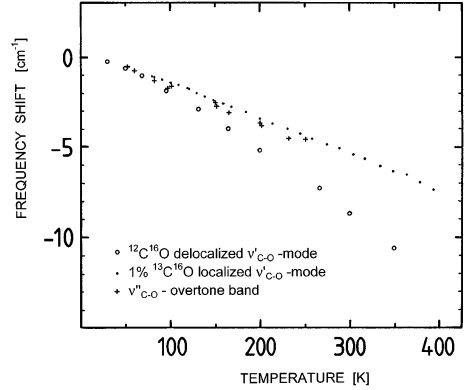}
\caption{\label{fig8-6-4}Frequency shifts of the vibrational bands with temperature:
crosses - overtone band, dots - fundamental of naturally abundant $^{13}$C$^{16}$O,
open circles - delocalized fundamental of $^{12}$C$^{16}$O.
Adapted from \cite{pj96}.}
\end{center}
\end{figure}
Experimental spectra at 30 K are shown in Fig.\ref{fig8-6-3}.
The one phonon mode frequency is at 2031 cm$^{-1}$. This has to be compared
to the naturally abundant $^{13}$C$^{16}$O frequency at 1941 cm$^{-1}$.
The corresponding blue shift for the adsorbate is thus due to additional stiffness
provided by the Ru surface coupling.
Excitation of two uncorrelated phonons would yield a two phonon continuum at about 4062 cm$^{-1}$.
The narrow line observed at 3940 cm$^{-1}$ can be thus attributed to a two-phonon bound state,
or a quantum discrete breather excitation.

The temperature dependence of the line positions also clearly shows,
that the two-phonon bound state line softens much slower than
the line of the one-phonon delocalized state (Fig.\ref{fig8-6-4}).
\begin{figure}[!t]
\begin{center}
\includegraphics[angle=0,width=0.6\textwidth]{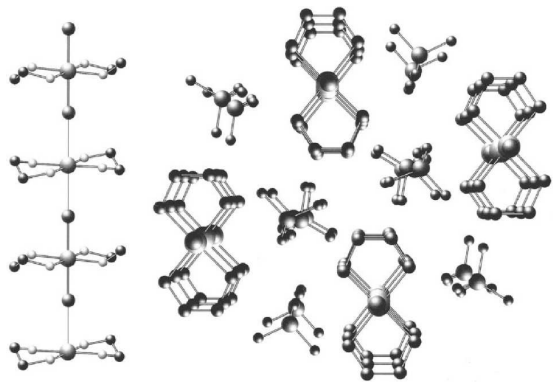}
\caption{\label{fig8-6-5}Structure of the PtCl crystal. One PtCl chain is shown on the left.
Each Pt atom is coordinated by two ethylenediamine units in a near
square planar geometry, while Cl ions connect pt sites along the chain.
The packing arrangement of the 1D chains and their ClO$_4^-$
counterions is shown on the right.
Adapted from \cite{sblssbws99prl}.}
\end{center}
\end{figure}
This is, among other facts, a strong indication that the observed red shift of the overtone
line is due to the formation of a localized two-phonon bound state, or a (quantum)
discrete breather.

\subsubsection{ In the bulk of solids} 

Vibrational spectra in the overtone or combination region of molecular crystals have been
studied intensively in the 1970s and 1980s. A pioneering theoretical proposal
was due to Agranovich, who predicted the existence of two-exciton bound states in various
molecular crystal materials \cite{vma70}. Experimental studies of infrared absorption spectra
for CO$_2$ crystals were conducted by Dows et al \cite{dadvs73} and gave evidence
of two-phonon bound states. Dressler et al studied the slow vibrational relaxation
of N$_2$, which also indicates the presence of many-phonon bound states \cite{kdoodas75}.
In a remarkable theoretical paper, Bogani calculated the spectrum of two phonon excitations
in molecular crystals \cite{fb78}, to some extent one of the first accurate
calculations of quantum discrete breathers.
More recently Bini et al reconsidered the theory of three-phonon bound states
in crystal CO$_2$ \cite{rbprsvshjj93}. While there certainly are many other 
results worth to be mentioned, we recommend reading related chapters in \cite{vmarmh83},\cite{aaonsekap01}.
\begin{figure}[!t]
\begin{center}
\includegraphics[angle=0,width=0.3\textwidth]{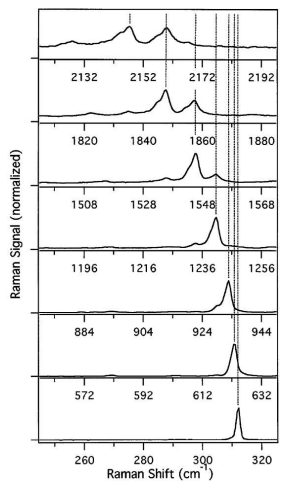}
\caption{\label{fig8-6-6}Fundamental and overtone spectra of isotopically pure Pt$^{35}$Cl.
Moving upward in each panel, each $x$ axis is offset by the appropriate
integral multiple of the 312 cm$^{-1}$ fundamental frequency. All spectra 
have been scaled vertically to equal peak intensities.
Adapted from \cite{sblssbws99prl}.}
\end{center}
\end{figure}
\begin{figure}[!t]
\begin{center}
\includegraphics[angle=0,width=0.5\textwidth]{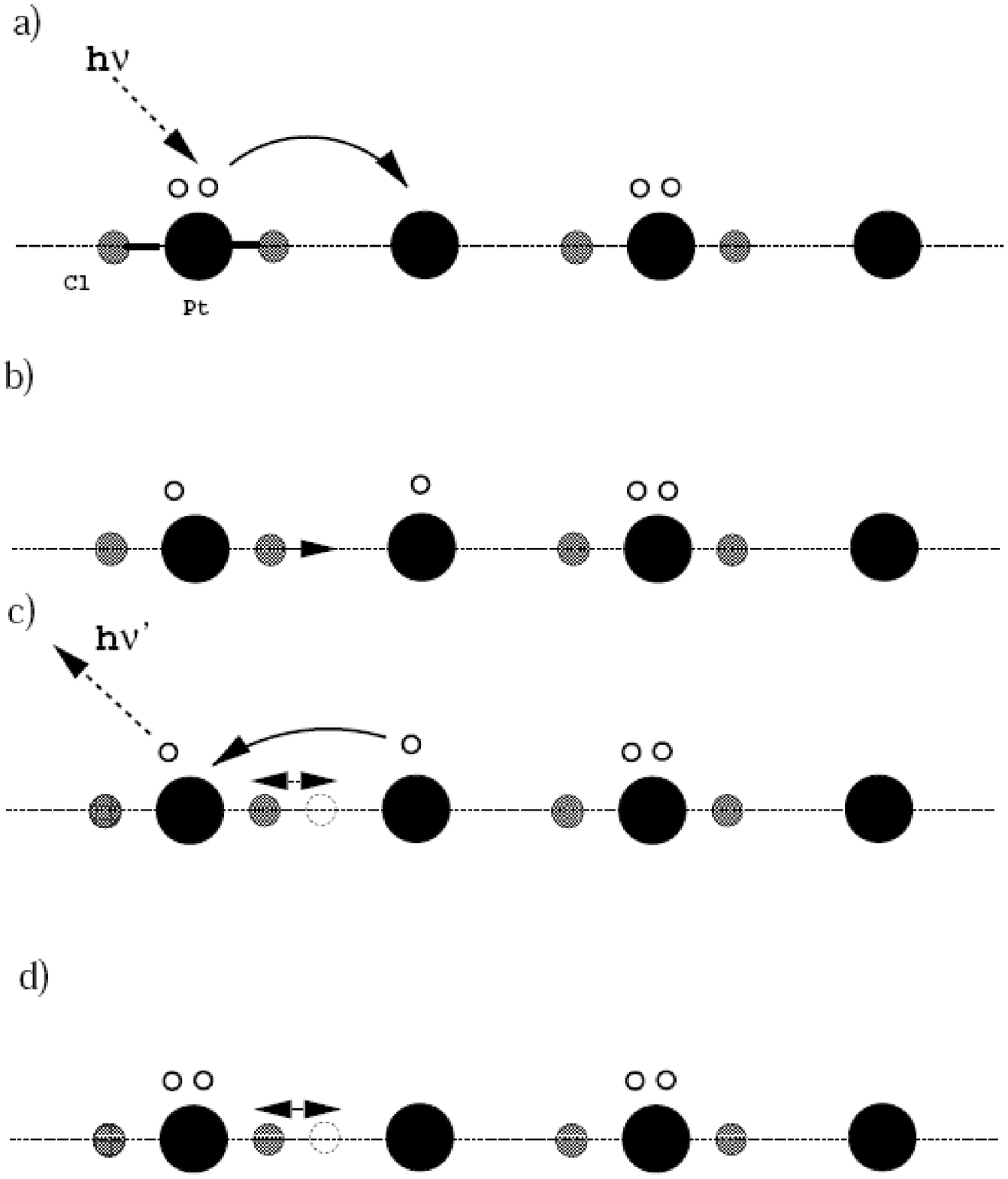}
\caption{\label{fig8-6-7}A simple picture of a resonant Raman scattering event in the localized
atomic limit. Large filled circles mark Pt ions, small grey circles mark
Cl ions. Open circles mark the positions of electrons.
Adapted from \cite{kkjmarb99}.}
\end{center}
\end{figure}

The pioneering studies of Swanson et al \cite{sblssbws99prl} 
have shown that up to seven phonons can bind and form a localized state.
The system of choice was a PtCl based crystal - a halide-bridged
mixed-valence transition metal complex, which is a model low-dimensional 
electronic material where the ground states can be systematically tuned
(with chemistry, doping, pressure, and temperature). It is a very strong charge-density wave (CDW)
example. The material is a well-formed crystal with a homogeneous lattice consisting
of quasi-1D chains (see Fig.\ref{fig8-6-5}). 
The CDW ground state consists of alternating Pt$^{+2}$ and Pt$^{+4}$
sites with a corresponding distortion of the chloride ions towards the Pt$^{+4}$ site.
Resonance Raman spectra were used to probe both ground and photoexcited states.
They probe the fundamental Cl-Pt-Cl stretch and the progression of many overtones.
At low temperatures, the fundamental exhibits a fine structure with up to six discrete,
well-resolved modes. The analysis of the evolution of the spectral structure in the
overtones was performed for isotopically pure samples, in order to avoid exciting
localized states due to isotopic disorder.
The fundamental and overtone spectra for the pure Pt$^{35}$Cl
sample are shown in Fig.\ref{fig8-6-6}. The data are presented in a 
stack plot in which each successive trace is offset along the horizontal
axis by increasing multiples of the fundamental frequency 312 cm$^{-1}$.
Such plots clearly expose the relation of features in the overtone
spectrum to multiples of the fundamental peak. The lowest energy
dominant feature in each trace (marked by vertical lines) demonstrates
a strongly increasing anharmonic redshift. Further, at higher overtones,
each of these dominant peaks recurs, offset by the fundamental frequency,
in the next trace above. A simple interpretation is that the lowest-energy dominant
peaks in the overtone spectra correspond to all quanta of vibrational energy
localized in approximately one PtCl$_2$ unit, while the higher energy peaks
correspond to having all quanta but one in a localized PtCl$_2$ unit combined
with one quanta in the more extended fundamental.
The schematic process of the energy transfer is shown in Fig.\ref{fig8-6-7}
and has been analyzed theoretically in \cite{kkjmarb99}.

An incoming photon at frequency $\nu$ is exciting an electron from a Pt$^{+2}$ to a
Pt$^{+4}$ site. The Cl ion between them starts oscillating. Finally the electron
relaxes back to its original position, and releases a photon with frequency $\nu '$.
The energy difference is remaining in a localized vibration.
The effect of isotope disorder was treated by Kalosakas et al \cite{gkarbaps02}.
The experimentally observed redshifts were also theoretically described
by Fehske et al \cite{hfgwhbarbmis00} using Peierls-Hubbard models, and
by Wellein et al using the Holstein model \cite{gwhf98}. 

Inelastic X-ray and neutron scattering was used by Manley et al 
\cite{memmyhs06,memmy+07,memjwlycghl08}
to probe phonon dispersion in $\alpha$-uranium single crystals.
Variation of temperature showed softening, and the abrupt appearance of
a new dynamical mode, without a typically observed phase transition.
The authors argue that this mode is a discrete breather, and forms due to
strong electron-phonon interaction.

Russell and Eilbeck reported evidence for moving breathers in the layered crystal muscovite
at 300 K \cite{fmrjce07}. Breathers were created by bombardment of the crystal
surface with heavy ions. Ejection of atoms at the opposite (shielded) crystal surface
was attributed to breathers, which were able to carry the vibrational energy
without dispersing over more than $10^7$ unit cells of the crystal.

Finally Abrasonis et al \cite{gawmxxm06} reported on anomalous bulk diffusion
of interstitial Nitrogen in steel microcrystals. N ions were deposited
in a micron-thick layer, and are trapped by local structures,
with a characteristic binding energy. Ar ion bombardment increases the N mobility
at depths far beyond the ion penetration depth. The authors see evidence
for coherent transfer of vibrational energy deep into the bulk of the material.

\section{Conclusions and outlook}

Progress in the understanding of classical discrete breathers evolved for two decades, and has significantly improved our understanding of the complexity in the dynamics of anharmonic lattice dynamics. The little input it needs to form discrete breathers - spatial discreteness and nonlinearity - were demonstrated to generate an impressive list of experimental observations in a large variety of different physical systems, with 
length and time scales ranging over many orders of magnitude. The quantization of the classical equations of motion
opened a whole new avenue of research on quantum breathers. For obvious computational reasons numerical studies
are much more involved in the quantum domain, restricting our potential of detailed understanding to either small lattices,
or few participating quanta. Nevertheless, a set of experiments, comparable in size to its classical counterpart,
has been reviewed where quantum breathers are detected. Still we anticipate this only to be the beginning of a much more
sophisticated undertaking. We may expect many new qualities to show up in the regime of large lattices and many participating quanta.
These limits are conceptually closing the gap to quantum interacting many body problems, which are at the heart
of condensed matter physics, and quantum computing. Yet another highly interesting, complicated and debated problem
is the combined influence of discreteness, nonlinearity, and disorder. Linear disordered wave equations allow for Anderson localization.
It is a thrilling modern theme to study the constructive and destructive interplay of all three ingredients.
The best is yet to come.

\end{document}